\documentclass[%
superscriptaddress,
 amsmath,amssymb,
 aps,showkeys,
floatfix,
]{revtex4-2}
\usepackage{amsmath}
\usepackage{graphicx}
\usepackage{epstopdf}
\usepackage[caption=false]{subfig}

\usepackage[breaklinks=true]{hyperref}
\usepackage{breakcites}
\usepackage{amssymb}
\usepackage{color}
\usepackage{enumerate}
\usepackage{setspace}
\usepackage{multirow}

\usepackage{booktabs,ragged2e,longtable}

\begin{document}

\title{Residential clustering and mobility of ethnic groups} 

\date{\today}

\author{Kunal Bhattacharya}
\email[Corresponding author: ]{kunal.bhattacharya@aalto.fi}
\affiliation {Department of Industrial Engineering and Management, Aalto University School of Science, Espoo, Finland}
\affiliation {Department of Computer Science, Aalto University School of Science, P.O. Box 15400, FI-00076 Aalto, Finland}
\author{Chandreyee Roy}
\affiliation {Department of Computer Science, Aalto University School of Science, P.O. Box 15400, FI-00076 Aalto, Finland}
\author{Tuomas Takko}
\affiliation {Department of Computer Science, Aalto University School of Science, P.O. Box 15400, FI-00076 Aalto, Finland}
\author{Anna Rotkirch}
\affiliation{Population Research Institute, V\"aest\"oliitto – Finnish Family Federation, Helsinki, Finland}
\author{Kimmo Kaski}
\affiliation {Department of Computer Science, Aalto University School of Science, P.O. Box 15400, FI-00076 Aalto, Finland}

\begin{abstract}
We studied residential clustering and mobility of ethnic minorities using a theoretical framework based on null models of spatial distributions and movements of populations. Using microdata from population registers we compared the patterns of clustering  amongst various socioethnic groups living in and around the capital region of Finland. Using the models we were able to connect the factors influencing intraurban migration to the spatial patterns that have been developed over time. We could also 
demonstrate the interrelationship of the movement and clustering with fertility. The observed clustering seems to be a combined effect of fertility and the tendency to migrate locally. The models also highlight the importance of factors like proximity to the city-centre, average neighbourhood income, and similarity of socioeconomic profiles.
\end{abstract}


\maketitle
\section{Introduction}

Quantifying the levels of residential segregation in urban settings is a challenging and complex problem. Several conceptual and methodological paradigms exist for studying socioeconomic heterogeneities observed in spatial distributions of populations. Important approaches include the formulation of spatial indices, recognizing the dimensions, cartographic projections,  modelling of the underlying dynamics, and the inclusion of demographic factors~\cite{theil1971note,white1983measurement,massey1988dimensions,dmowska2017comprehensive,clark1991residential,bayer2004drives}. A major fraction of these studies have included populations within cities in the United States and major European countries especially the ones with a long history of immigration and settlement~\cite{zubrinsky2003dynamics,clark2015multiscalar,braamaa2008dynamics,musterd2017socioeconomic}. Both from the viewpoint of research and policy making, 
segregation is largely considered as a negative outcome at the collective level even when resulting from unintended choices of individuals~\cite{murie2004social,andersson2007mix}. 

In this work, we propose to quantify segregation using the general ideas of scaling the laws of which at urban level 
have been widely used to characterize how macrolevel quantities of cities, such as GDP and productivity, vary with population size~\cite{bettencourt2007growth}. The types of scaling relations have been argued to be generally different depending upon whether the quantity is related to social interactions or to human engineered infrastructures~\cite{bettencourt2013origins,barthelemy2019statistical}. Recent studies have also examined temporal scaling laws in which a relevant quantity for a city is studied as a function of its growing population and, therefore, quantifies the historical evolution~\cite{depersin2018global,keuschnigg2019scaling,bettencourt2020interpretation}. Apart from such allometric scaling, the notions of fractality and power-laws have been long-investigated in urban systems. The latter approach has also been used to model the  dynamical nature of city growth including intra- and inter-urban migration~\cite{barthelemy2019statistical,verbavatz2020growth,reia2022modeling}. 

We begin with a model for the propensity of ethnic minority groups to inhabit spatially contiguous areas. We modify a null model~\cite{louf2016patterns} for spatial distribution of populations by assuming that the number of individuals of a group in an area would scale as the group population in the neighbouring areas. In effect, our model provides a measure of spatial clustering that has been conceived as one of the dimensions of segregation~\cite{massey1988dimensions,reardon2004measures}. In the model, we additionally include a set of socioeconomic variables that could independently influence the spatial clustering of minorities. Secondly, we use similar models for studying migration flows of groups to understand how spatial clustering could be reinforced over time. We use micro data from Finland to compare the spatial and temporal clustering of different ethnic groups residing in the capital region. 

\begin{figure}[htp]

\includegraphics[height=7cm]{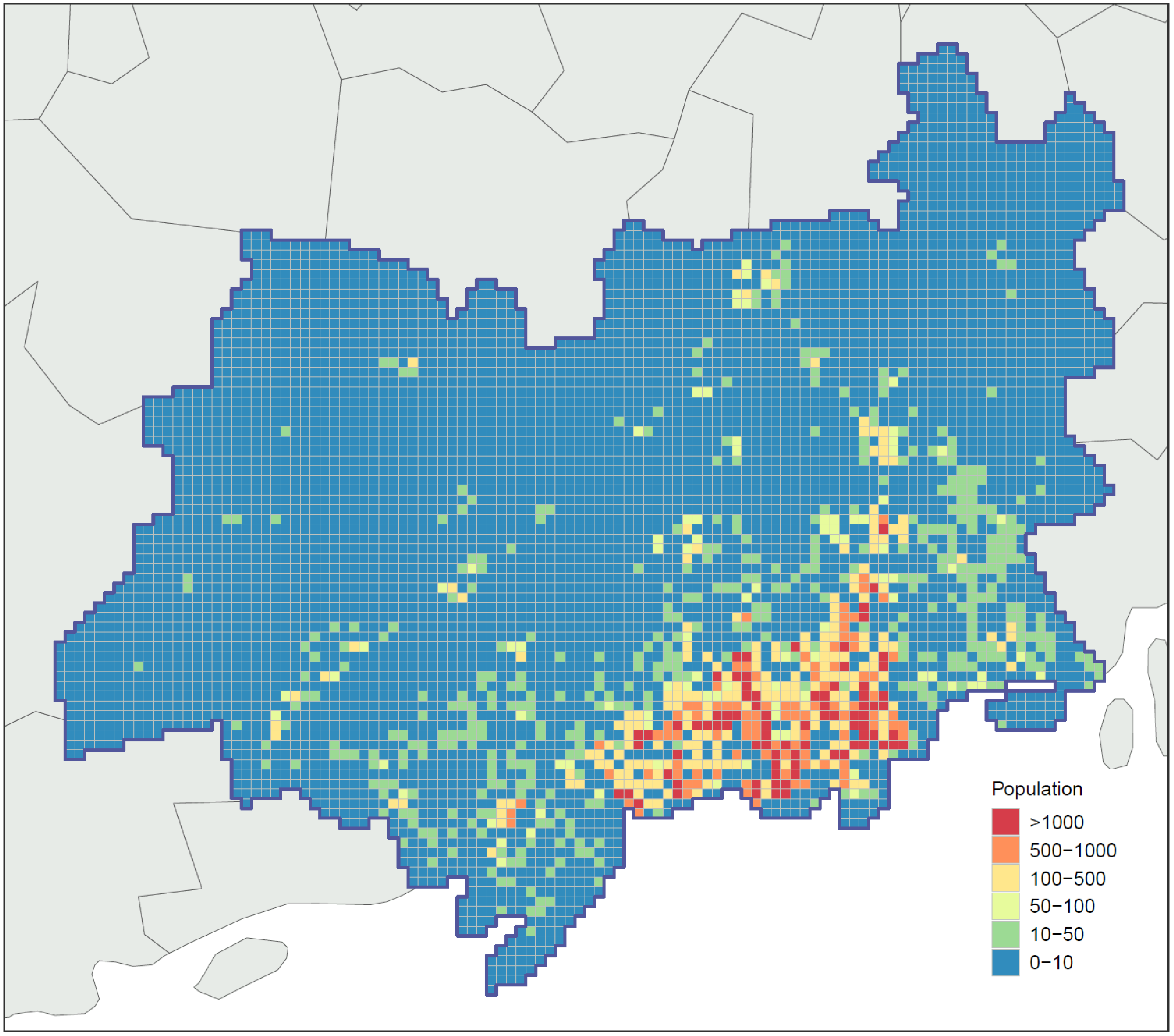}\\

\bigskip

\includegraphics[width=\textwidth]{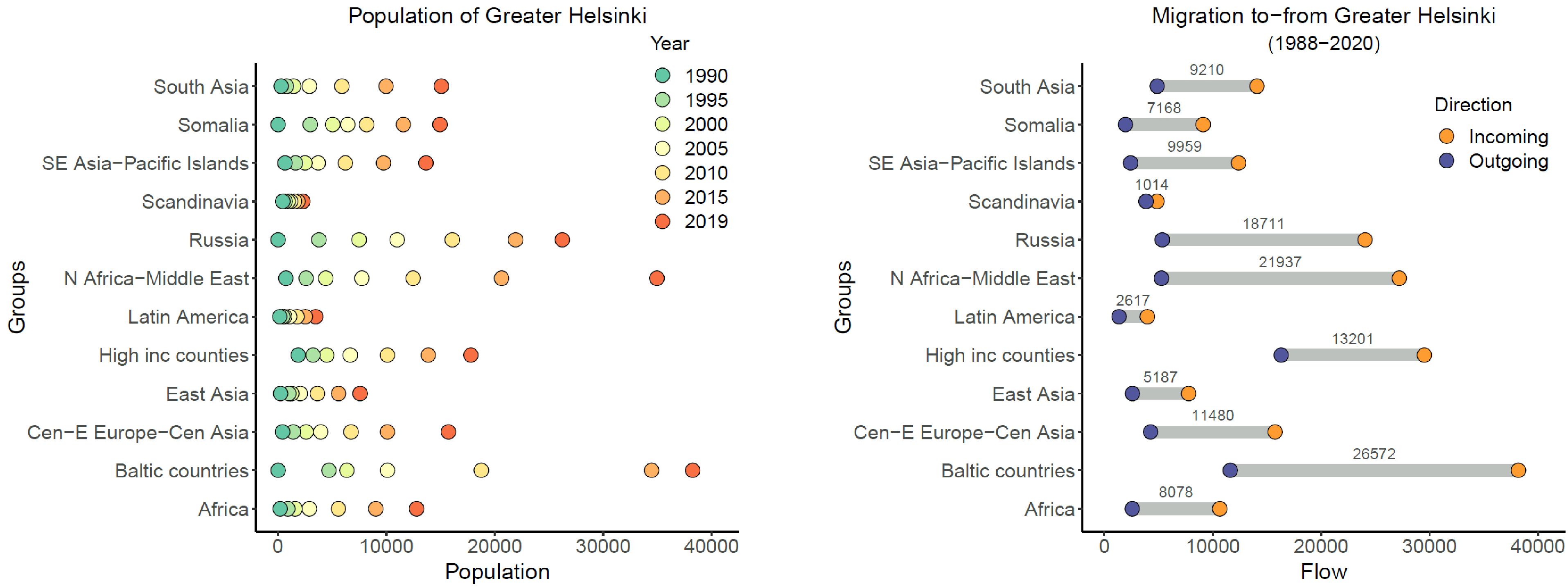}

\caption{(Top) The region of Greater Helsinki (GH) which includes the capital city of Finland and sixteen other municipal regions. The municipalities surrounding  GH are shown with grey borders. The grid is constituted by 1 km$^2$ cells from which statistical data is available. The population is predominantly composed of native Finnish speakers. The heat map corresponds to the population of different socioethnic groups in 2019 including the migrants as well as the Swedish speakers native to Finland. (Bottom left) The population of different groups in GH are shown corresponding to different years. Africa denotes nations from the sub-Saharan part. (Bottom right) The aggregated migration flows during the period of 1988-2020 for the groups. The flows include international migration as well as population entering  from (or leaving for) other regions of Finland. The net flows (difference between the total incoming and total outgoing flows) are denoted in grey. For the native Finnish and Swedish speakers: In 1990, there were roughly 80 thousand native Swedish speakers in GH, and by 2019 the population marginally dropped. In this period, the population of the native Finnish speakers grew by 20$\%$ to around 1.2 million. For both these groups the flows in either directions aggregated over the period were rather balanced - around 0.6 million for Finnish speakers and 30 thousand Swedish speakers.}
\label{fig_intro}
\end{figure}

Compared to North America and other European countries, and given the fact that Finland became an independent nation in 1917, the history of migration in Finland can be considered rather recent~\cite{vaattovaara2010contextualising}. The primary sources of migration to Finland have been the European countries specially, Sweden and Russia. Additionally, immigration took place through refugee waves at different points of 
time. The topic of social segregation of ethnic minorities has been well studied for the Finnish population owing to the availability of microlevel data population registers. Past research included descriptive studies, measurement of segregation indices, integration efforts, and impact of policies related to job markets and housing, among the other aspects~\cite{vaattovaara2010contextualising,zhukov2015has,kauppinen2019unravelling,ansala2022immigration}. Notably, the municipalities in the capital region including Helsinki and neighbouring areas are known to have official strong desegregation policies~\cite{dhalmann2009housing,dhalmann2013explaining}. By the end of 2019, around 8$\%$ of the total Finnish population was composed of people with foreign background, out of which around one-half resided in the Greater Helsinki metropolitan region, which includes the capital area and the surrounding municipalities. Also, foreign language speakers accounted for over 70$\%$ of the net migration into the region~\cite{statfi}. In Fig.~\ref{fig_intro} we show the population of different socioethnic groups (see Sec.~\ref{data} for details) inside Greater Helsinki that were considered in the study.

\section{Results}


\subsection{Spatial clustering}

As a starting point for a model to quantify clustering we began with the framework presented by~\citet{louf2016patterns}. The authors considered a null model for the distribution of the population of different groups over the area units in a city that is unsegregated. Adopting the framework to the current problem, let $n_{i,g}$ and $n_{i}$ denote the population belonging to a group $g$ and the total population in an area unit $i$ of the Greater Helsinki (GH) region measured at any point of time, respectively. Here, $g$ represents the groups enumerated in the section~\ref{data} as well as the group of Finnish-speaking Finns who in this dataset comprise the majority of the population (see the SI for population of each group). 
Given that the total number in group $g$ is $N_g$, such that $\sum_{g}N_{g}=N_{\text{GH}}=\sum_{i}n_{i}$, where $N_\text{GH}$ is the total population in GH, and when $n_{i,g}<<n_{i}$, the null model would predict the following expectation value for populations,
\begin{equation}
    \text{E}\left[ n_{i,g}\right]= N_{g}\frac{n_{i}}{N_\text{GH}}.
\label{null_model}
\end{equation}
However, in reality, several socioeconomic factors are expected to influence the distribution of the population of 
different groups. The propensity to cluster based on ethnicity is one such factor, which we would like to measure. We assume that the population of a group in a location $i$ could be predicted by the population of people from similar ethnicity in the adjoining areas. We quantify this effect by considering that the expectation value in the above model is modified by a prefactor $\widetilde{n}_{i,g}^{\beta_s}$, where $\widetilde{n}_{i,g}$ is the population of individuals belonging to group $g$ per unit area in a neighbourhood $\mathcal{N}_i$ and the exponent $\beta_s$  is the measure of 
clustering. To measure $\beta_s$ for different groups we formulated this as separate regression model for each group as follows
\begin{equation}
\log n_{i,g}=\beta_0 + \beta_{N}\log N_{g} +  \beta_n \log n_i + \beta_{s}\log \widetilde{n}_{i,g} + \epsilon_{i,g},
\label{base_model}
\end{equation}
where $\epsilon_{i,g}$ is a normally distributed error with zero mean. To calculate
$\widetilde{n}_{i,g}$, we chose $\mathcal{N}_i$ as the set of eight units in the Moore neighbourhood around the area unit $i$.  We refer to Eq.~\ref{base_model} as the `base model'.

Although, we started from a null model by~\citet{louf2016patterns}, the variable $n_i$ can be alternately thought of as a proxy for factors that 
would typically influence the growth of population at the $i$-th location. Previously,~\citet{jones2015modelling,jones2015ethnic} presented a similar approach for 
structuring models around the expectation number for counts within the categories in the contexts of assimilation and segregation~\cite{jones2015modelling,jones2015ethnic}. Studies using different modelling schemes have also considered measurement of segregation that is conditional or controlled with respect to different observed characteristics of groups that 
simultaneously influence the clustering of populations~\cite{aaslund2009measure,aaslund2010will}. Our approach is similar to the earlier works that have included additional explanatory variables in the models to account for the characteristics of individuals, groups, and locations~\cite{kalter2000measuring,bayer2004drives,andersson2014workplace}. 

Within the current scheme, we add the following three variables:
\begin{enumerate}[(a)]
    \item Income of populations is a factor understood to be intricately related to the dynamics of residential segregation in cities as detailed by, for example,~\citet{bayer2004drives}. The mean income of individuals at a location could be representative of the housing prices. However, income of individuals and groups, in general, can both be a cause and an outcome of the sorting process of groups into different locations~\cite{king1973racial,reardon2011income}. Here, we include $\langle m\rangle_i$, the average income in an area unit $i$ that was calculated from summing of the yearly disposable incomes of all the individuals in the unit irrespective of group affiliations. By including this variable in the model we expect $\beta_{s}$ to capture the aspect of clustering that is unrelated to income~\cite{bayer2004drives,harsman2006ethnic}.
    
    \item In their work~\citet{massey1988dimensions} introduced ``centralization'' as one of the dimensions of segregation to explain possible tendency in an immigrant population to reside near the centre of a city that 
    was tacitly assumed to be the hub of economic activity~\cite{duncan1955methodological}. We take this possibility into consideration via the variable $d_{\text{cen},i}$, the distance between an area unit $i$ and the centre of the GH area. Although the GH area is generally not considered to be mono-centric, the central location does have substantially larger share of the population and employments~\cite{vasanen2012functional,granqvist2019polycentricity}.  
    
    \item Previous studies have discussed linkages between the residential and workplace segregation, and generally it is understood that individuals who are employed in the same workplace are prone to also reside near each other, and vice versa~\cite{bayer2004drives,andersson2014workplace,glitz2014ethnic}. Therefore, we expect people with similar skills to also cluster in terms of residential locations irrespective of the their ethnic affiliations. We take this aspect into account by including the variable $D_\text{KL}(P_{g,i}||Q_i)$, which is the Kullback-Leibler (KL) distance measured between the empirical distributions of socioeconomic statuses $P_{g,i}$ and $Q_i$. The dataset contains information on the socioeconomic statuses of individuals, for instance, whether a person is self-employed or an upper-level employee (see SI Fig.~\ref{fig-appendix-6} for the complete list). Using this information we calculated $P_{g,i}$ for people belonging to a group $g$ residing at location $i$. Similarly, we calculated $Q_i$ taking into account all the people residing in $i$ and in the neighbourhood $\mathcal{N}_i$. A small KL distance implies a high similarity between the socioeconomic profiles of people from group $g$ and the rest of the population residing in and around the area, and therefore, could independently explain a large value of $n_{i,g}$.

\end{enumerate}   

With the inclusion of the above variables the `full model' reads as follows: 
\begin{equation}
\begin{split}
\log n_{i,g}= &\beta_{0} + \beta_{N,g}\log N_{g} +  \beta_{n,g} \log n_i + \beta_{s,g}\log \widetilde{n}_{i,g}+\beta_{m,g} \log \langle m\rangle_i \\
&\quad  +\beta_{\text{cen},g} \log d_{\text{cen},i} + \beta_{\text{KL},g} \log (1+D_\text{KL})+ \epsilon_{i,g}.
\label{full_model}
\end{split}
\end{equation}
Here, we did not control for variables that have direct functional dependence on age, gender and education of the residents, which were  
considered in some of the previous studies~\cite{bayer2004drives,aaslund2009measure,glitz2014ethnic,aaslund2010will}. We assumed that the possible variation in $n_{i,g}$ due to the latter variables were accounted for by the income and the socioeconomic status. Additionally, we assumed that there were no omitted variables that were significantly correlated with $\log n_i$, in particular. We used ordinary least squares (OLS) to evaluate the coefficients in the model using data from the year 2019. There is, however, a level of complexity involved in estimating the coefficient of  $\widetilde{n}_{i,g}$. See the Methods (\ref{sec:estimation_spatial}) for the details and the relative fit indices for the null, the base, and the full model.

\begin{figure}[t]
\includegraphics[width=\textwidth]{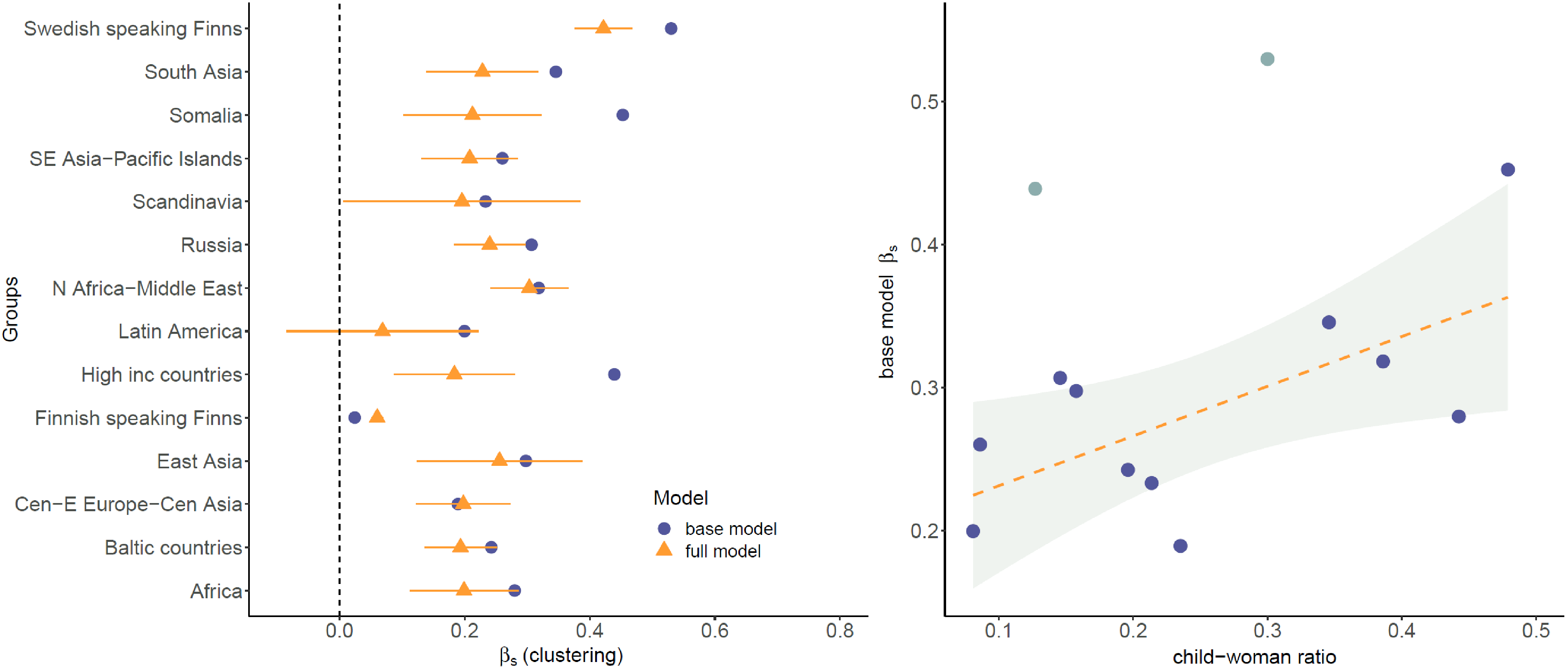}
\caption{(Left) The measure of clustering is plotted for the different groups. In general, the introduction of additional explanatory variables deflates the clustering as can be observed by comparing the results from the base and the full models. The error bars for the full model depict 95$\%$ confidence intervals. The bars for the base model are omitted for clarity. (Right) The clustering is plotted against the child-woman ratio which is a cross-sectional measure of fertility of groups. Considering the groups -- Swedish speaking Finns and high income countries as outliers, the dashed line has slope of  0.35 ($p<.05$, $r^2=0.43$). This plot depicts that a part of the variability in clustering could be explained by the differences in fertility.}
\label{fig1}
\end{figure}



The coefficient quantifying spatial clustering (Fig.~\ref{fig1}) obtained from the full model containing all the explanatory variables was generally lower than the one obtained from the more restricted base the model. The reductions were substantial especially for the populations corresponding to high income-countries and Somalia. The coefficient was found to a range between zero and 0.5, and was highest for the native Swedish speaking 
population. The spatial clustering as measured using the base model could be a reflection of two distinct processes taking place over longer periods of time, fertility and migration. As a measure of fertility for the cross-sectional data we used the child-woman ratio (CWR) for the different groups. It was calculated as the ratio between the number of children under five and the number of women aged between 15 and 49, considering the individuals in the population register during 2019. In general, the clustering was found to increase with the fertility of the groups (Fig.~\ref{fig1}-right). Given two groups with the same population sizes, the one with the higher CWR would tend to have larger families, and therefore would show a higher clustering in space. In the next section we 
will investigate the aspect of migration. 

\begin{figure}[t]
\centering
\includegraphics[width=16.cm]{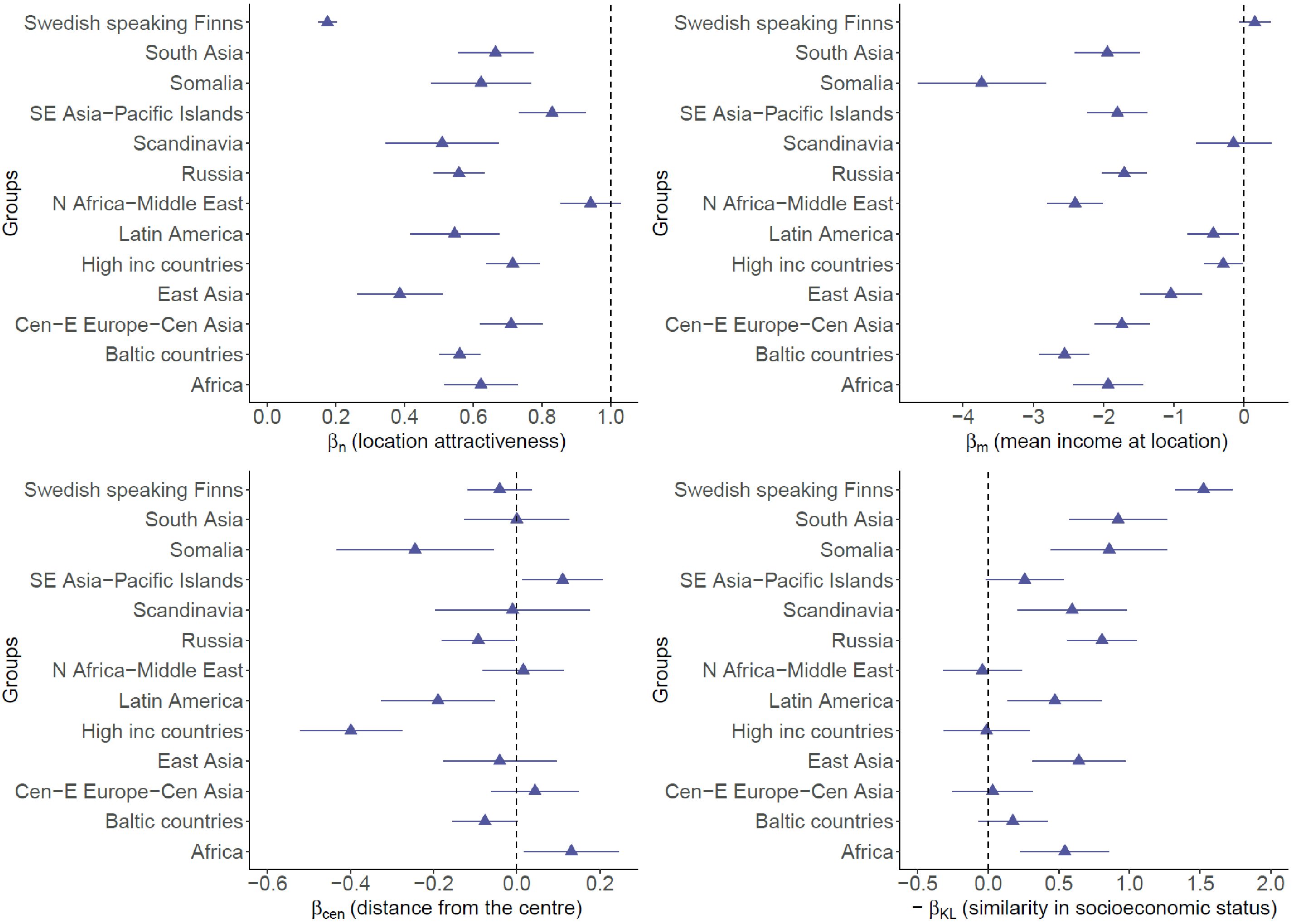}
\caption{The coefficients in the full spatial model corresponding to the variables which were assumed to independently predict the population of the different groups. The relevance of the variables within the model are mentioned inside the brackets.  
}
\label{fig2}
\end{figure}

Next, we show the coefficients (exponents) corresponding to the other explanatory variables. In Fig.~\ref{fig2} (top-left) we show the dependence on the total population in an area unit, which we have interpreted as the general attractiveness of an area. For the groups, in general, $\beta_{n}>0$ which implies that the higher the total population in an area unit, the higher is the group's population. For some groups like  North Africa-Middle East the dependence was the strongest while it was weakest for the Swedish speaking Finns.  The top-right shows $\beta_{m}\leq0$, which implies inverse relationship between the group population and the mean income
at the locations. While this effect was pronounced for a group like Somalia,  for the groups like high income-countries, Swedish speaking Finns, Scandinavia, and Latin  America the effect was almost absent. The left-bottom also shows an inverse relation with the distance from the centre of Helsinki. The latter relationship appeared to be weak as $\beta_\text{cen}$ was non-significant in many of the cases. For the high income-countries the population appeared to decrease fastest with the distance from the centre. In the bottom right we show $-\beta_\text{KL}$, which quantifies the tendency of individuals to reside beside other individuals of similar socioeconomic status.   

\subsection{Migration and clustering}

We compared the changes in population for the different groups in the different area units due to migration, in particular, the movements internal to the Greater Helsinki. Although, the overall population growth of ethnic minorities could be largely attributed to migration from outside GH, the local changes are dominated by the movements between parts of the metropolitan area. This was evidenced by comparing the magnitudes of the intra-region flows and the other types of flows (combining emigration, immigration, and flows from or to other regions of Finland), and examining the ratio between those two quantities (see SI Fig.~\ref{fig-appendix-15}). Similar to the case of the population distribution, we assumed a null model for the flow of a group at any area unit $i$:
\begin{equation}
    \text{E}\left[F_{i,g}^\text{(M)}\right]= F_{g}^\text{(M)} \frac{F_{i}^\text{(M)}}{F_{\text{GH}}^\text{(M)}},
    \label{null_model-flow}
\end{equation}
where $M$ denotes whether the flow is incoming or outgoing, $F_{i,g}$ is the inflow or outflow in the area unit $i$ for group $g$, $F_{i}$ is the total flow in area unit $i$, $F_{g}$ is the total flow for the group $g$,  and $F_{\text{GH}}$ is the flow for the entire region.  Note, that the condition $\sum_i F_{i}=F_\text{GH}=\sum_{g} F_{g}$ holds separately for inflow and outflow. 
In addition, 
the condition $F_\text{GH}^{\text{in}}$=$F_\text{GH}^{\text{out}}$ was approximately satisfied depending on the filtering and the 
quality of data.

For the inflows we modified the null model in the following way. First, we used the additional variable $\log n_{i,g}$ to quantify the `presence' of a group within the area unit.  Second, as the migration was intra-region we considered the possible presence of social gravity whereby people would tend to relocate to nearby
locations~\cite{barthelemy2019statistical}. Ideally, a model on gravity would imply studying the flows between pairs of locations, but in the current model involving 
just focal locations, we included the total outflow $\widetilde{F}_{i,g}^\text{(out)}$ of the group from the neighbourhood $\mathcal{N}_i$ as an explanatory variable. Also note that 
in case of a gravity model we would need to deal with noisier and sparser group-level flows resulting in poorer model fits. Therefore, the inflow model in terms of aggregated flows was specified as,
\begin{equation}
\log \mathcal{F}_{i,g;\tau,\Delta}^\text{(in)}=\alpha_{0,g}^\text{(in)} +\alpha_{F,g}^\text{(in)}\log \mathcal{F}_{i;\tau,\Delta}^\text{(in)} + \alpha_{p,g}^\text{(in)}\log n_{i,g;\tau}+ \alpha_{s,g}^\text{(in)}\log \widetilde{\mathcal{F}}_{i,g;\tau,\Delta}^\text{(out)}+\epsilon^\text{(in)}_{i,g},
\label{base_model-inflow}
\end{equation}
where $\log n_{i,g,\tau}$ is the population of group $g$ in the year $\tau$, and the $\mathcal{F}$'s are flows aggregated over 
the period between $\tau$ and $\tau+\Delta$. The latter aggregation was done to reduce the noise in the yearly flows $F$. The coefficient $\alpha_{F}$ is expected to account for the population-level factors influencing the flows to an area unit, and $\alpha_{p}$ is to account for the presence of the group $g$. The coefficient $\alpha_{s}$ should quantify the tendency to relocate between nearby locations. For the outflows emanating from an area unit we solely considered the attributes of the focal unit:
\begin{equation}
\log \mathcal{F}_{i,g;\tau,\Delta}^\text{(out)}=\alpha_{0,g}^\text{(out)} +\alpha_{F,g}^\text{(out)}\log \mathcal{F}_{i;\tau,\Delta}^\text{(out)} + \alpha_{p,g}^\text{(out)}\log n_{i,g;\tau}+\epsilon^\text{(out)}_{i,g}.
\label{base_model-outflow}
\end{equation}

The Eqs.~\ref{base_model-inflow} and~\ref{base_model-outflow} constituted as our base models for the intraregion migrations. Further, we extended these models by including the centralization variable $d_{\text{cen},i}$ and the average income in the area unit in the year $\tau$, $\langle m\rangle_{i;\tau}$ as explanatory variables. For brevity, we provide only the full model corresponding to the inflow: 
   \begin{equation}
\begin{split}
      \log \mathcal{F}_{i,g;\tau,\Delta}^\text{(in)}=&\alpha_{0,g}^\text{(in)}+\alpha_{F,g}^\text{(in)}\log \mathcal{F}_{i;\tau,\Delta}^\text{(in)} + \alpha_{p,g}^\text{(in)}\log n_{i,g;\tau}+ \alpha_{s,g}^\text{(in)}\log \widetilde{\mathcal{F}}_{i,g;\tau,\Delta}^\text{(out)}\\
    &\quad +\alpha_{\text{cen},g}^\text{(in)}\log d_{\text{cen},i} + \alpha_{m,g}^\text{(in)}\log \langle m\rangle_{i;\tau}  +\epsilon^\text{(in)}_{i,g}.
    \label{full_model-inflow}  
\end{split}
\end{equation}

The aggregation of flows was done by summing over the flows $F_{\tau'}$ in each year $\tau'$ which gave $\mathcal{F}_{\tau,\Delta}= \sum_{\tau'=\tau+1}^{\tau+\Delta}F_{\tau'}$, where $\tau\in\{1995,2000,2005,2010,2015\}$ and $\Delta=5$. While it was possible to run the regressions separately for each $\tau$, we mean-centred the variables for each $\tau$ and then performed the regressions after pooling the data  from the different time windows. The mean-centering was done to take into account the population growth that would have taken place over time in the area units. Ten out of the thirteen groups that we investigated (not including the native Finnish speakers) had adjusted-$R^2$ for the inflow models in the range 0.4--0.8. The lower values for the rest three (for example, Scandinavians) could have resulted from smaller sample sizes. For the outflow model, in general, we found it to have a higher explanatory power but with marginal difference between the base and the full model (see SI Fig.~\ref{fig-appendix-8}).   Finally, we combined the inflow and outflow models to arrive at the following general expression for the ratio of the inflow to the outflow across area units:
\begin{equation}
    \frac{\mathcal{F}_{i,g}^\text{(in)}}{\mathcal{F}_{i,g}^\text{(out)}}\sim \widetilde{\mathcal{F}}_{i,g}^{\text{(out)}\alpha_s}\,n_{i,g}^{\varDelta\alpha_p} \,\langle m\rangle_{i}^{\varDelta\alpha_m} \,d_{\text{cen},i}^{\varDelta\alpha_\text{cen}},
    \label{inflow_outflow_ratio}
\end{equation}
where, $\varDelta\alpha_p =\alpha_p ^\text{(in)}-\alpha_p^\text{(out)}$, $\varDelta\alpha_m =\alpha_m ^\text{(in)}-\alpha_m^\text{(out)}$, and $\varDelta\alpha_\text{cen} =\alpha_\text{cen} ^\text{(in)}-\alpha_\text{cen}^\text{(out)}$. 

\begin{figure}[t]
\centering
\includegraphics[width=0.85\linewidth]{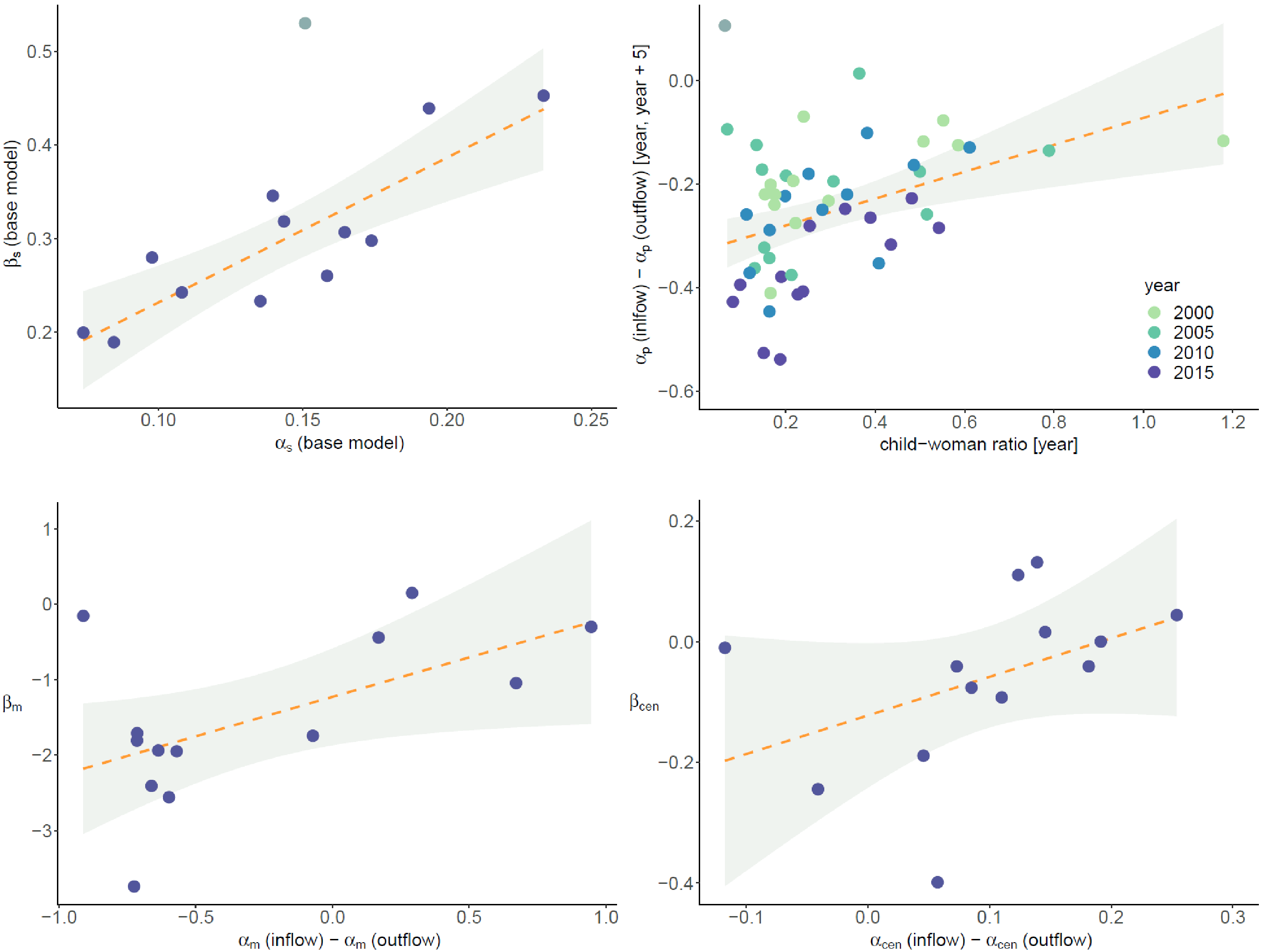}
\caption{(Top-left) Relationship between clustering ($\beta_s$) measured in the spatial base model and the coefficient (exponent) of neighbourhood outflows ($\alpha_s$) in the inflow base model. The point corresponding to the Swedish speaking Finns is considered as an outlier to a linear relationship and is shown in a lighter shade. The slope of the fitted line is 1.54 ($p<.001$, $r^2=0.74$). The shaded region is the 95$\%$ confidence interval. We show $\beta_s$'s joint dependence on $\alpha_s$ and the CWR in Tab.~\ref{table-1}. (Top-right)  The coefficient of group population, $\varDelta\alpha_p =\alpha_p ^\text{(in)}-\alpha_p^\text{(out)}$ from the model of inflow-outflow ratio (Eq.~\ref{inflow_outflow_ratio}) is plotted against the child-woman ratio. Each data point represents the regression coefficient corresponding to a group and a specific period over which the flows were aggregated. While the CWR was calculated for the years ($\tau$) as shown in the legend, the flows corresponded to $[\tau,\tau+\Delta]$ with  $\Delta=5$. 
The dashed line has slope of  0.26 ($p<.01$, $r^2=0.18$).  Note, in general, for all the groups  $\alpha_p^\text{(out)}>\alpha_p^\text{(in)}$. A single data point (indicated in grey) was considered as an outlier. The two plots at the bottom show the behaviour of the coefficients from the spatial full model  against relevant coefficients in the inflow-outflow ratio. (Bottom-left) Coefficient of the mean income at area unit $\beta_m$ versus $\varDelta\alpha_m =\alpha_m ^\text{(in)}-\alpha_m^\text{(out)}$ (slope = 1.04, $p<.05$, $r^2=0.32$). (Bottom-right) Coefficient of distance from the city centre $\beta_\text{cen}$ versus $\varDelta\alpha_\text{cen} =\alpha_\text{cen} ^\text{(in)}-\alpha_\text{cen}^\text{(out)}$ (slope = 0.64, $p>.1$, $r^2=0.18$).}
\label{fig3}
\end{figure}

We investigated whether $\alpha_s$, which captures the tendency to move locally could be associated to the the observed clustering in space (Fig.~\ref{fig3}). First, considered only the base models in which the variables were primarily geometric in nature. 
Treating the observation for Swedish speaking Finns as an outlier we were able to observe a significant association (slope) between $\beta_s$ and $\alpha_s$. 
Given that $\beta_s$ was separately found to depend on the child-woman ratio, we examined $\beta_s$'s joint dependence on $\alpha_s$ and CWR (Tab.~\ref{table-1}). With the base models we found $\beta_s$ to significantly vary with both CWR and $\alpha_s$ (adj. $R^2$=0.88). In case of the full models only the relationship to $\alpha_s$ remained significant (adj. $R^2$=0.73). 

In the premise of the base models we also checked for any association between the coefficient $\varDelta\alpha_p$ and the CWR (Fig~\ref{fig3} top-right). Naively, the observed behaviour  could indicate that the groups with higher fertility accumulate population from other areas over time. However, the vertical axis shows that for the groups $\varDelta\alpha_p<0$ implying 
that the inflow to outflow ratio varies inversely with the local group population (density). Therefore, 
individuals irrespective of group affiliation would tend to migrate out over time from areas with higher concentration. Apparently, the rate at which the latter occurs is influenced by the overall CWR, which 
is likely to impact the inflow and outflow rates in contrasting ways. While a higher CWR would mean that the inflow volume is constituted by the movement of larger families,  the outflow from an area could generally be impeded by the presence of larger number of children within the local population. Interestingly, the observed relationship between $\varDelta\alpha_p$ and the CWR appeared to be more robust in comparison to the individual variations with  $\alpha_p ^\text{(in)}$ and $\alpha_p^\text{(out)}$ (see SI Fig.~\ref{fig-appendix-17}). The correlations of the CWR with  $\alpha_p ^\text{(in)}$, $\alpha_p^\text{(out)}$, and $\varDelta\alpha_p$ were 0.28, -0.05, and 0.44, respectively. The result $\varDelta\alpha_p<0$, however,  reflects the dominant pattern, and fluctuations in time windows and in area units are a possibility whereby the inflow of a certain group would surpass its outflow. For the native Finnish speaking population we found both the $\alpha_p$'s to be negligibly small and $\varDelta\alpha_p=-0.02$ for a fit on the entire period.  Being the majority population, the null model stated in Eq.~\ref{null_model-flow} is almost an identity, in turn making the coefficients-$\alpha_F$ almost unity and diminishing the covariation of the flows with other variables.

\begingroup
\squeezetable
\setlength{\tabcolsep}{18pt}
\renewcommand{\arraystretch}{1.5}
\begin{table}[t]
\scriptsize
\caption {Interrelationships between the coefficients from the spatial model, migration model(s), and the CWR, quantified using linear fits.} 
\begin{tabular}{l r l}
\midrule

  \multicolumn{1}{c}{model coefficients} &  \multicolumn{1}{c}{slope(s)} &  \multicolumn{1}{c}{model types}      
            
	                                                 \\    \midrule

 $\varDelta\alpha_p$, CWR & 0.26$^{**}$ & \multirow{2}{*}{base models}  \\ \cmidrule{1-2}                                               
 \multirow{2}{*}{$\beta_s$, (CWR, $\alpha_s$)}   &  (0.23$^{**}$, 1.16$^{***}$)$^\text{1}$    \\ 
 \cmidrule{2-3}
 
  &  (0.03, 0.91$^{***}$)$^\text{2}$                                     & \multirow{3}{*}{full models}        \\

          \cmidrule{1-2}
 {$\beta_m$}, $\varDelta\alpha_m$   & 1.04$^{*}$ \\  \cmidrule{1-2}
{$\beta_\text{cen}$}, $\varDelta\alpha_\text{cen}$   & 0.64 & \\
\midrule
\end{tabular}
\begin{tabbing}
$^{\mathrm{1}}$ Groups SF and HI  were excluded (see~Fig.~\ref{fig1}).\\
$^{\mathrm{2}}$ Group SF was excluded.\\
$p$ values are indicated as $^{*}p < .05$; $^{**}p < .01$; $^{***}p < .001$.
\end{tabbing}
\label{table-1}
\end{table}
\endgroup

Using the full models we also examined the relations between the income-at-location and the distance-to-centre coefficients. The coefficient $\varDelta\alpha_m$ took both positive and negative values in a range between -1.0 and 1.0, which would imply contrasting behaviour in terms of the inflow to outflow ratio (Fig~\ref{fig3} bottom-left). For some groups the overall migration over time has been directed towards areas with high average income, while for some the opposite occurred. The relationship between $\beta_m$ and $\varDelta\alpha_m$ was found to be significant with an approximate unit slope. The relationship between $\beta_\text{cen}$ and $\varDelta\alpha_\text{cen}$ was found to be qualitatively similar (Fig~\ref{fig3} bottom-right). Here the positive values of $\varDelta\alpha_\text{cen}$ would imply the concentration building up over time towards the periphery of the city, and negative values the converse. However, the overall slope although positive was not found to be statistically significant.

%


\section{Discussion}
We analysed spatial clustering and mobility of socioethnic groups using a modelling approach based on scaling relations built on the top of null models. First, the models were able to establish the relevance of different factors, which separately govern the population distribution of the groups. Second, the framework could empirically relate the coefficients (exponents) characterising the clustering in space to the coefficients characterising the incoming and outgoing population flows from area units inside the region. Third, our results indicate that the clustering in space is likely to be reinforced by the migration at shorter distances as well as the presence of large sized families in the population. Fourth, all groups, in general, were found to move out of the regions of 
higher concentrations with a rate negatively dependent on fertility. The third and fourth, taken together is indicative of a diffusive process for the spread of population in contiguous areas.  Our findings are consistent with previous studies conducted using different measures of segregation, especially in the context of Helsinki~\cite{kauppinen2019unravelling}. These studies have shown that the mobility has, generally, led 
to the dilution of immigrant neighbourhoods~\cite{zwiers2018trajectories} and how higher fertilities have led 
to higher concentration of ethnic minorities~\cite{finney2009population}. Overall, the framework could identify the differences between the groups as well as reveal
a set of `stylized facts' which in turn could be valid more or less universally in different urban settings.

The clustering measured within the base models, conditional solely on populations or flows, could still be interpreted geometrically. This was strictly not possible when socioeconomic variables like the average neighbourhood income were included in the models. Importantly, the association between the clustering ($\beta_s$) from the spatial model and the coefficient of outflows from neighbourhood ($\alpha_s$) of 
the migration model was not weakened  when additional variables like, the average yearly income at area unit ($\langle m\rangle_{i}$) were included. This 
underscores the importance of the neighbourhood migration in generating clustered population distributions to different degrees across groups.

The child-woman ratio with which significant association was found using the base models, however, was found to loose relevance when the full models were studied.  We considered the child-woman ratio to be reflective of larger cohabiting families that could strengthen the concentration of members from the same group. However, larger number of children would also imply lower average income at the area level. Therefore, the inclusion of the $\langle m\rangle_{i}$ in the full models could have removed the variation with the child-woman ratio. Note, that the correlation between the average income of groups at the population level ($\langle m\rangle_g$) and the base model-$\beta_s$ was -0.72 ($p<.05$, after excluding the groups SF and HI). But when tested alongside  $\alpha_s$ and CWR, the variable $\langle m\rangle_g$ did not show any significant association or yield 
additional explanatory power. Expectedly, for the full spatial model (excluding the group SF), the correlation of $\langle m\rangle_g$ with $\beta_s$ became insignificant (-0.45, $p>.1$) but was present in the case of $\beta_m$ (0.84, $p<.001$).

 It was also interesting to note that for the Swedish speaking Finns the clustering remained severely underpredicted by neighbourhood migration or the child-woman ratio. Swedish speaking Finns are 
 a historical part of the Finnish society, and their concentration in one of the municipalities in the studied region is over thirty percent. Therefore, unlike other groups their access and choices of residences would be guided by multiple additional factors. The Figs.~\ref{fig1} and~\ref{fig2} showed that the residential pattern simultaneously had the highest spatial clustering in terms of ethnicity and, the highest level of similarity of socioeconomic status within neighbourhoods. This was likely due to the movement and reorganisation processes over longer periods of time within the areas already maintaining ethnic concentration of Swedish speakers. Such a pattern would not be visible for groups that were late entrants to the society.


\section{Methods}

\subsection{Data}
\label{data}
We used anonymized register data from Statistics Finland that included separate yearly longitudinal modules on basic demographic variables of individuals, their locations, and information on migration events during the years 1987-2020. For our research we focused on the Greater Helsinki metropolitan area that consists of 17 municipal regions. For assigning area units to the individuals we utilized the coarse-grained information on location in the EUREF-FIN coordinate system (ERTS89-TM35FIN). This allowed assigning residences of individuals to 1 km $\times$ 1 km square areas in a grid. The information is usually available to researchers if there are more than three residents per area unit. For modelling the spatial clustering we primarily used the datasets from 2019, and for growth and migration we used the mobility and location sets from 1995 onward.  

The ethnicities of the individuals were based on a `country of ethnicity' which we assigned by combining multiple types of data. First, we checked whether an individual has a Finnish or non-Finnish background, and additionally whether the person was born abroad or in Finland.
We selected the individuals with non-Finnish background and then utilized the records on migration as detailed in section~\ref{algorithm}. 
From a combination of basic demographic and location data files we ended up with 4088 area units in the GH area that were populated on the average during period of investigation. In particular, for the year 2019, the basic module contained records of 5.52 million residents from the entire Finland out of those we could locate 1.57 million to GH. Out of the latter there were around 233 thousand individuals who had foreign backgrounds with 81$\%$ being born outside Finland. Among all individuals with foreign background we could assign a unique country of origin to around 191 thousand individuals.  In studying the growth of populations during the period spanning 1995-2020 we could assign countries to a total of around 490 thousand individuals. In addition, our study included Swedish speaking Finns who are considered native to Finland alongside native Finnish speakers. In 2019, there were around 83 thousand such individuals in GH who could be directly identified by their language available as a part of their basic information. In total, this makes around 20$\%$ percent people in GH to be either of non-natives or non-Finnish speakers. Note, that the percentage of population in Finland with immigrant background is generally lower than other countries in the Nordic region or Europe.

\subsection{Algorithm for assigning ethnicities}
\label{algorithm}

The ethnicities of the individuals with foreign background included in the present work were assigned using their respective migration information from Statistics Finland. The data included the year of migration, type of migration (immigration or emigration), country of departure/arrival, first and second citizenship along with their country of birth.  
 For an individual of non-Finnish background but born in Finland a country of ethnicity (COE) was assigned according to the citizenship at the time of migration giving preference to the first citizenship. For those who were born abroad, the COE was assigned using the information on immigration. For individuals with multiple immigration, we considered only the first year of immigration into Finland. Individuals with multiple migrations have been removed from the study since they include multiple countries and the COE cannot be determined in such a case. 
The pseudocode followed for assigning the COE is given in SI Fig.~\ref{fig-appendix-14}.  

Next, using the information on the COE we assigned the individuals with foreign background to groups following a slightly modified version of super-regions used in the Global Burden of Diseases, Injuries, and Risk Factors Study (GBD) 2017~\cite{murray2018population}. The groups  following GBD correspond to the different regions of the world, namely, Central Europe-Central Asia, East Asia, High-Income Countries, Latin America, North Africa-Middle East, Africa (sub-Saharan), South East Asia-Pacific Islands, and South Asia. Additionally, given Finland's past and more recent history of immigration we considered the groups: Russia, Somalia, Baltic countries, and Scandinavia. We also considered the Swedish-speaking population from Finland which is a group having Finnish heritage but is generally considered as an ethnic minority. See SI for the details on the groups.

\subsection{Estimation of spatial model coefficients}
\label{sec:estimation_spatial}

We primarily used OLS to extract the coefficients for the spatial model. This estimation of  $\beta_{s,g}$ may suffer from endogeneity bias as $\widetilde{n}_{i,g}$ contains in itself the dependent variable. Therefore, we compared the OLS estimates with those from a spatial autoregressive (SAR) lag  framework~\cite{lesage2009introduction} which is known to properly circumvent spatial dependencies. Also, the SAR lag specification has an extended interpretation in terms of a steady-state description encompassing feedback and reinforcement effects occurring over longer periods of time between neighbouring regions. The latter aspect addresses the simultaneity issue which arises from the fact that all the variables were measured in the same year.

To apply the SAR lag method we slightly reformulated Eq.~\ref{full_model} whereby $\widetilde{n}_{i,g}$ is expressed as a linear combination of spatial lags of the dependent variable:
\begin{equation}
\begin{split}
\log n_{i,g}= &\beta_{0,g} +  \beta_{n,g}\log n_i + \beta_{s,g} \sum_{j\in \mathcal{N}_i} w_{ij}\log n_{j,g}\\
&\quad + \beta_{m,g} \log \langle m\rangle_i + \beta_{\text{cen},g}  \log d_{\text{cen},i} + \beta_{\text{KL},g} \log (1+D_\text{KL})+ \epsilon_{i,g},
\label{sar_model}
\end{split}
\end{equation}
where $w_{ij}$ is a weight for an adjacent unit $j$. Given $z_i$ as the number of neighbouring areas of $i$ that have non-zero population of group $g$, we chose $w_{ij}=1/z$ when $\log n_{j,g}>0$ and $w_{ij}=0$, otherwise (see below for the filtering conditions).  The SAR lag model was estimated for the individual groups where $N_{g}$ was absorbed inside $\beta_{0,g}$. To estimate the parameter we used the R-package \texttt{spatialreg} which first performs an optimization to calculate $\beta_{s,g}$ and then uses generalized least squares for the other coefficients~\cite{bivand2015comparing}. The  overall agreement between the OLS and the SAR models was high with the Pearson correlation corresponding to the regression coefficients $\beta_s$, $\beta_n$, $\beta_m$, $\beta_\text{cen}$, and $\beta_\text{KL}$ being 0.77, 0.99, 0.99, 0.97,  and 0.98, respectively (see Fig.~\ref{fig-appendix-16} for a comparison between the $\beta_s$'s for different the groups). Note, the replacing of the logarithm of the neighbourhood population in Eq.~\ref{full_model} with  the sum of logarithms in Eq.~\ref{sar_model} would account for some differences. 

Additionally, we employed a hierarchical linear regression model~\cite{jones2015modelling,jones2015ethnic} which allowed us to compare models beginning  from the null to the full model encompassing data for all the groups (see Tab.~\ref{table-s3} for the different fit indices)~\cite{bates2015fitting}. The overall fit quality was improved with the addition of the explanatory variables as the pseudo-$R^2$~\cite{nakagawa2017coefficient} increased by 10$\%$ at each stage. For OLS the adjusted-$R^2$ for the null, the base, and the full model lied in ranges 0.23--0.56, 0.33--0.75, and 0.44--0.77, respectively (see Fig.~\ref{fig-appendix-5}).

In our analysis we used the following filtering conditions:  $n_{i,g}<n_i$ and $n_{i,g}\geq N_\text{min}$, where  $N_\text{min}$ is a fixed threshold for $n_{i,g}$. The first condition was a weaker version of the requirement $n_{i,g}<<n_i$ for the null model (Eq.~\ref{null_model}) to be valid. The second condition overall reduced the noise in the distributions of the explanatory variables, as well as for the distributions $P_{g,i}$ and $Q_i$ used for calculating the KL distance. In Eq.~\ref{sar_model} we used $\log n_{i,g}:=0$ for $n_{i,g}<N_\text{min}$. All our results correspond to $N_\text{min}=10$ and we have examined $\beta_s$'s and $\beta_n$'s sensitivities to $N_\text{min}$ in the SI (see Fig.~\ref{fig-appendix-13}). We used a similar condition $n_{i,g;\tau}\geq N_\text{min}$ for the migration models as well.

\section{Acknowledgements}
This work was (partly) supported by NordForsk through the funding to The Network Dynamics of Ethnic Integration, project number 105147.

%

\clearpage
\setcounter{figure}{0}
\setcounter{section}{0}
\setcounter{table}{0}
\renewcommand{\thesection}{S\arabic{section}}  
\renewcommand{\thetable}{S\arabic{table}}  
\renewcommand{\thefigure}{S\arabic{figure}}

\section*{Supporting Information}

\begingroup
\setstretch{1.0}
\setlength{\tabcolsep}{10pt} 
\begin{longtable}[h]{p{4.5 cm}p{8 cm}}
\caption{Grouping of countries following a modified version
of super-regions used in the Global Burden of Diseases, Injuries, and Risk Factors Study
(GBD) 2017.}\\
    

\hline
\hline
\textbf{Group} & \textbf{Country of ethnicity}  \\ 
\hline
\hline

\endfirsthead
\caption{(Continued) Grouping of countries.}\\
 \hline
 \hline
\textbf{Group} & \textbf{Country of ethnicity}  \\ 
\hline
\hline
\endhead
\hline
\endfoot
\hline\hline
\endlastfoot
 {\scriptsize Africa (AF), sub-Saharan} & {\scriptsize Angola, Benin, Botswana, Burkina Faso, Burundi, Cameroon, Cape Verde, Central African Republic, Chad, Comoros, Congo, Ivory Coast, Saint Helena, Djibouti, DRC, Equatorial Guinea, Eritrea, Eswatini, Ethiopia, Gabon, Gambia, Ghana, Guinea, Guinea Bissau, Kenya, Lesotho, Liberia, Madagascar, Malawi, Mali, Mauritiana, Mozambique, Namibia, Niger, Rwanda, Sao Tom\'e \& Principe, Senegal, Sierra Leone, South Africa, South Sudan}\\

{\scriptsize Baltic countries (BL)} & {\scriptsize Latvia, Lithuania, Estonia}\\

{\scriptsize Central + Eastern Europe \& Central Asia (CEA)} & {\scriptsize Armenia, Azerbaijan, Georgia, Kazakhstan, Kyrgyzstan, Mongolia, Tajikstan, Turkmenistan, Uzbekistan, Albania, Bosnia-Herzegovania, Croatia, Montenegro, North Macedonia, Serbia, Slovenia, Yugoslavia, Bulgaria, Czech Republic, Hungary, Poland, Romania, Slovakia, Belarus, Moldova, Ukraine}\\

{\scriptsize East Asia  (EA)} & {\scriptsize China, Hong Kong, Macau, North Korea, Taiwan}\\

{\scriptsize High income countries (HI)} & {\scriptsize Japan, Singapore, South Korea, Brunei, Andorra, Austria, Belgium, Cyprus, France, Germany, Greece, Ireland, Israel, Italy, Liechtenstein, Luxembourg, Malta, Netherlands, Portugal, San Marino, Spain, Switzerland, United Kingdom, Australia, Canada, Newzealand, United States of America}\\

{\scriptsize Latin America (LA)} & {\scriptsize Antigua-Barbados, Bahamas, Barbados, Belize, Bermuda, Bolivia, Cuba, Dominica, Dominican Republic, Ecuador, Grenada, Guyana, Haiti, Jamaica, Peru, Puerto Rico, Saint Lucia, St. Kitts \& Nevis, St. Vincent \& Grenadines, Saint Martin, Suriname, Trinidad-Tobago, Virgin Islands, Brazil, Colombia, Costa Rica, El Salvador, Guatemala, Honduras, Mexico, Nicaragua, Panama, Paraguay, Venezuela, Argentina, Chile, Uruguay}\\

{\scriptsize North Africa \& Middle East (NM)} & {\scriptsize Afghanistan, Algeria, Bahrain, Egypt, Iran, Iraq, Jordan, Kuwait, Lebanon, Libya, Morocco, Oman, Palestine, Qatar, Saudi Arabia, Sudan, Syria, Tunisia, Turkiye, United Arab Emirates, Yemen}\\

{\scriptsize Russia (RU)} & {\scriptsize Russia}\\

{\scriptsize Scandinavia (SC)} & {\scriptsize Denmark, Greenland, Iceland, Norway, Sweden}\\
{\scriptsize South-East Asia \& Pacific Islands (SEA)} & {\scriptsize Cambodia, East Timor, Indonesia, Laos, Malaysia, Maldives,
 Mauritius, Myanmar, Philippines, Seychelles, Thailand, Vietnam, American Samoa, Federated States of Micronesia, Fiji, Guam,
Kiribati, Marshall Islands, Nauru, Northern Mariana Islands, Palau, Papua-New Guinea, Samoa, Solomon Islands, Tonga, Tuvalu, Vanuatu}\\
{\scriptsize Somalia (SO)} & {\scriptsize Somalia}\\
{\scriptsize South Asia (SA)} & {\scriptsize Bangladesh, Bhutan, India, Nepal, Pakistan, Sri Lanka}
\\
{\scriptsize Swedish speaking Finns (SF)} & {\scriptsize Finland} \\
{\scriptsize Finnish speaking Finns (FF)} & {\scriptsize Finland} 



\label{table-s1}
\end{longtable}

\endgroup

\begin{figure}[h]
\centering
\caption{The following algorithm used to assign a country of ethnicity (COE) to individuals with foreign background.}
\fbox{\includegraphics[width=0.75\textwidth]{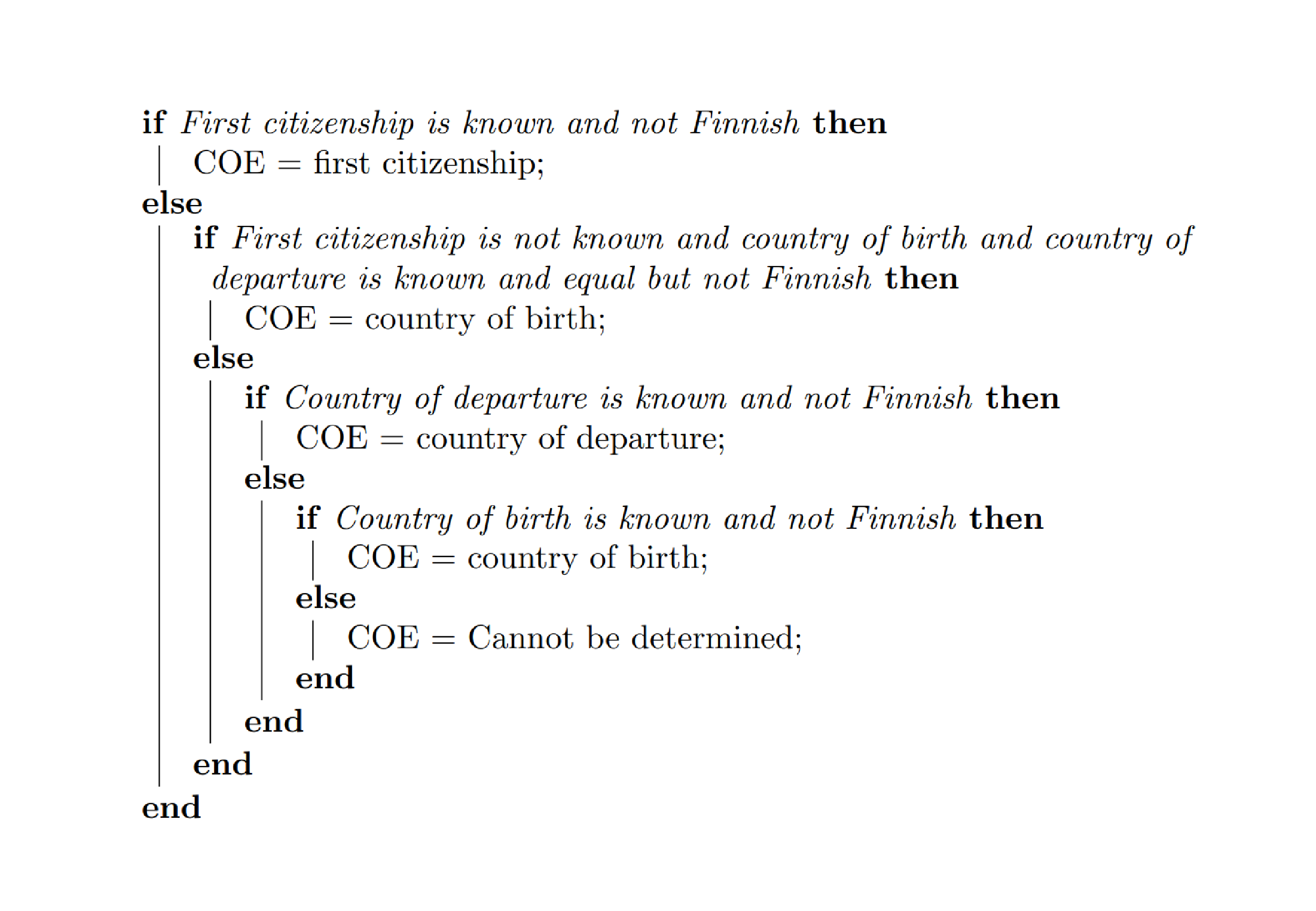}}
   
\label{fig-appendix-14}
\end{figure}

\begingroup
\squeezetable
\setlength{\tabcolsep}{18pt}
\renewcommand{\arraystretch}{1.5}
\begin{table}[h]
\scriptsize
\caption {Area units and population for each group in the final sample for spatial analysis corresponding to the year 2019.} 
\begin{ruledtabular}
\begin{tabular}{rlrr}
\textbf{No.}& \textbf{Group} & \textbf{Area units} & \textbf{Group population} \\     \hline\hline

1&  {\scriptsize Africa sub-Saharan} & {\scriptsize 270} & {\scriptsize 11,882}\\

2&{\scriptsize Baltic countries } & {\scriptsize 562} & {\scriptsize 35,795}\\

3& {\scriptsize Central + Eastern Europe \& Central Asia } & {\scriptsize 335}& {\scriptsize 14,283}\\

4&{\scriptsize East Asia  } & {\scriptsize 201} & {\scriptsize 6377}\\

5& {\scriptsize High income countries } & {\scriptsize 345} & {\scriptsize 15,959}\\

6&{\scriptsize Latin America } & {\scriptsize 116} & {\scriptsize 2236}\\

7&{\scriptsize North Africa \& Middle East} & {\scriptsize 408} & {\scriptsize 33,959}\\

8&{\scriptsize Russia } & {\scriptsize 433} & {\scriptsize 24,724}\\

9&{\scriptsize Scandinavia } & {\scriptsize 52} & {\scriptsize 933}\\
10&{\scriptsize South-East Asia \& Pacific Islands } & {\scriptsize 290} & {\scriptsize 12,030}\\
11&{\scriptsize Somalia } & {\scriptsize 201} & {\scriptsize 14,375}\\
12&{\scriptsize South Asia } & {\scriptsize 286} & {\scriptsize 14,248}
\\
13&{\scriptsize Swedish Speaking Finns} & {\scriptsize 931}  & {\scriptsize 77,966}
\\
14&{\scriptsize Finnish Speaking Finns} & {\scriptsize 2270}  & {\scriptsize 1,208,010}
\end{tabular}
\end{ruledtabular}
\label{table-s2}
\end{table}
\endgroup


\begin{figure}[h]
\centering
\includegraphics[width=0.75\textwidth]{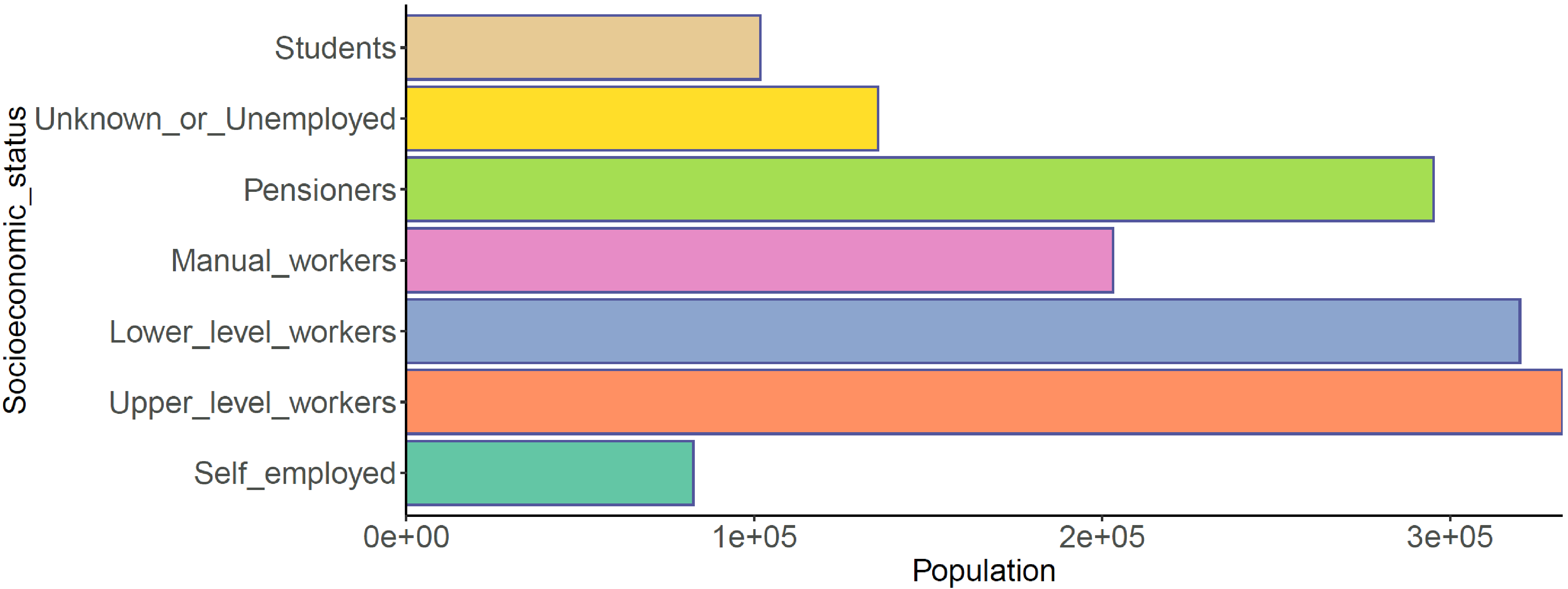}
\caption{Different categories of socioeconomic status and the distribution of the foreign background population in these categories. (The native Finnish and Swedish speaking population are not counted.)   
}
\label{fig-appendix-6}
\end{figure}

\begin{figure}[h]
\centering
\includegraphics[width=0.7\textwidth]{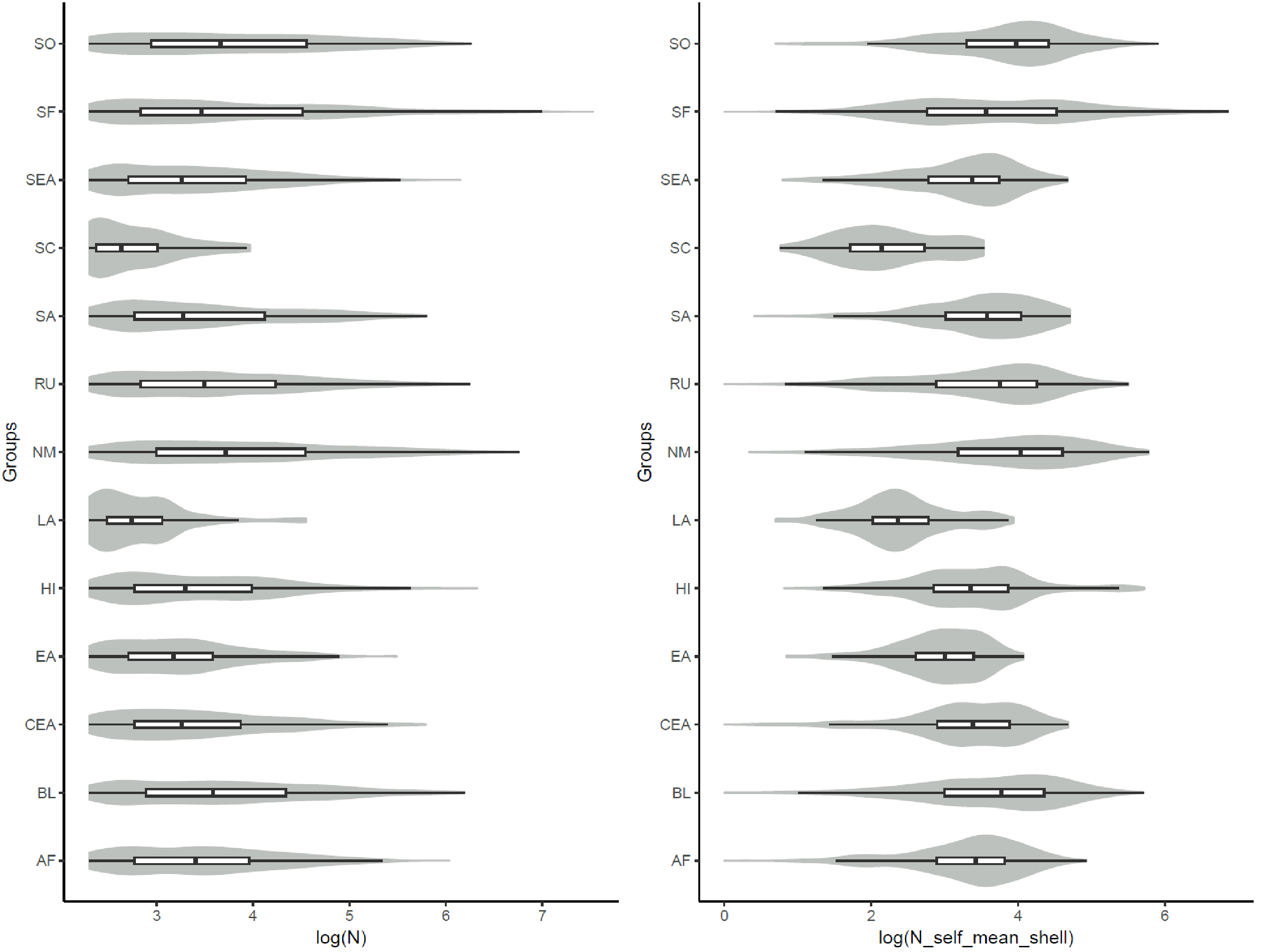}
\caption{Descriptive statistics for the variables, $\log n_{i,g}$ (left) and $\log\widetilde{n}_{i,g}$ (right). The plots show smoothed (mirrored) distributions of the variables in grey. The boxes show the interquartile range with the median. The horizontal lines on either sides denote spans of $\pm1.5$ interquartile range. 
}
\label{fig-appendix-1}
\end{figure}

\begin{figure}[h]
\centering
\includegraphics[width=0.7\textwidth]{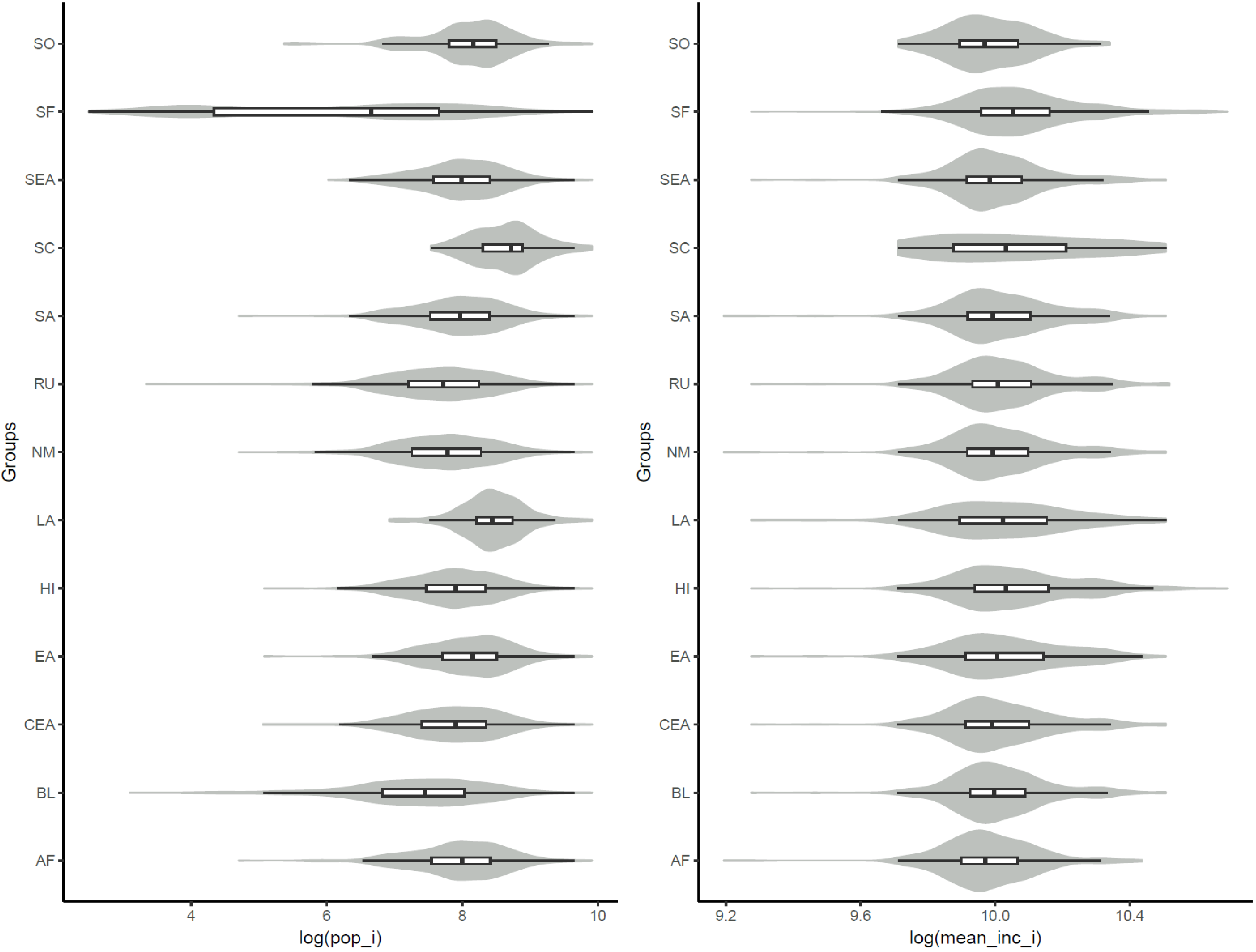}
\caption{Descriptive statistics for the variables, $\log n_{i}$ (left) and $\log \langle m\rangle_i$ (right). The plots show smoothed (mirrored) distributions of the variables in grey. The boxes show the interquartile range with the median. The horizontal lines on either sides denote spans of $\pm1.5$ interquartile range.  
}
\label{fig-appendix-2}
\end{figure}

\begin{figure}[h]
\centering
\includegraphics[width=0.7\textwidth]{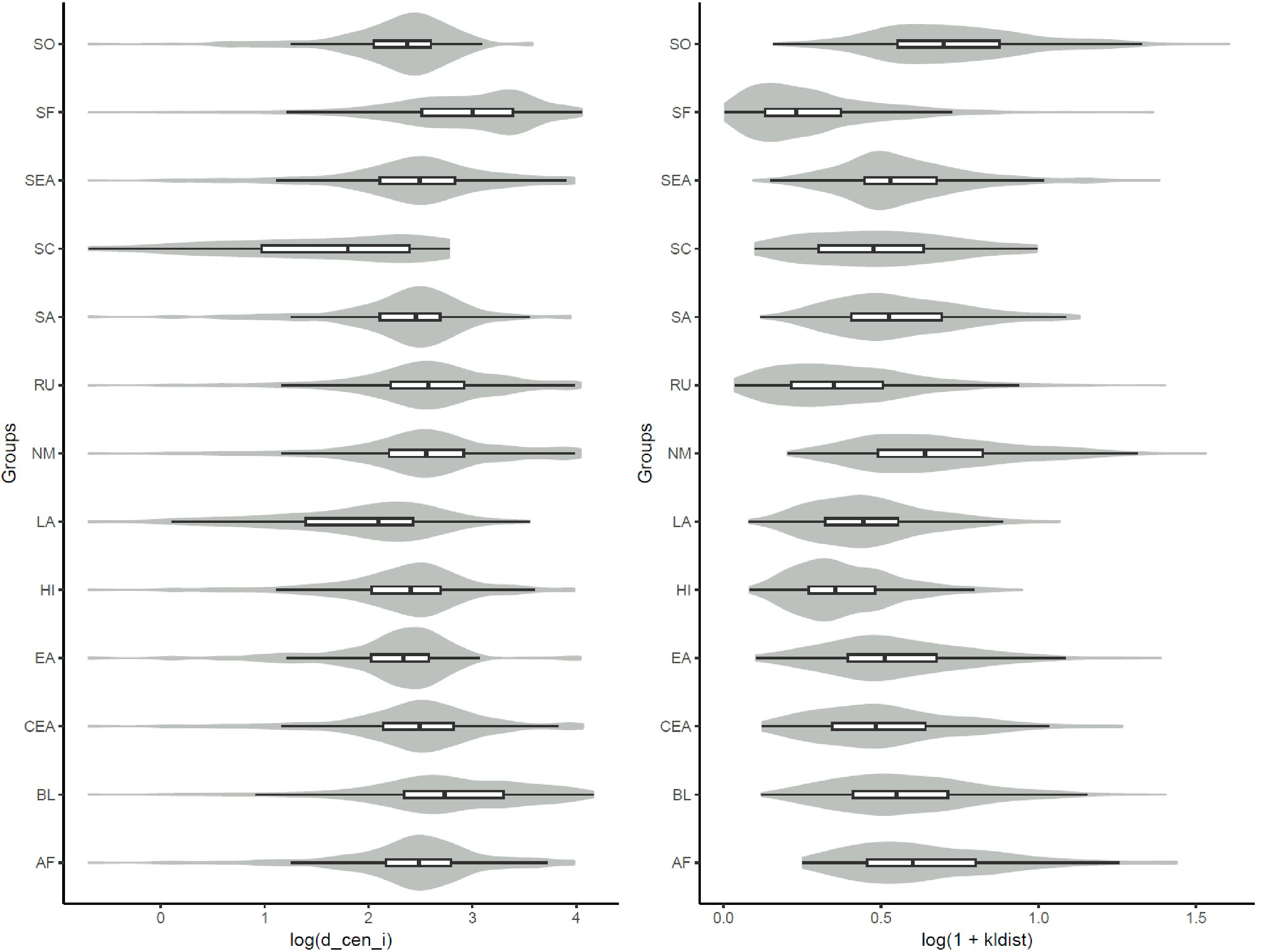}
\caption{Descriptive statistics for the variables, $\log d_{\text{cen},i}$ (left) and $\log (1+D_\text{KL})$ (right). The plots show smoothed (mirrored) distributions of the variables in grey. The boxes show the interquartile range with the median. The horizontal lines on either sides denote spans of $\pm1.5$ interquartile range.  
}
\label{fig-appendix-3}
\end{figure}

\begin{figure}[h]
\centering
\includegraphics[width=0.7\textwidth]{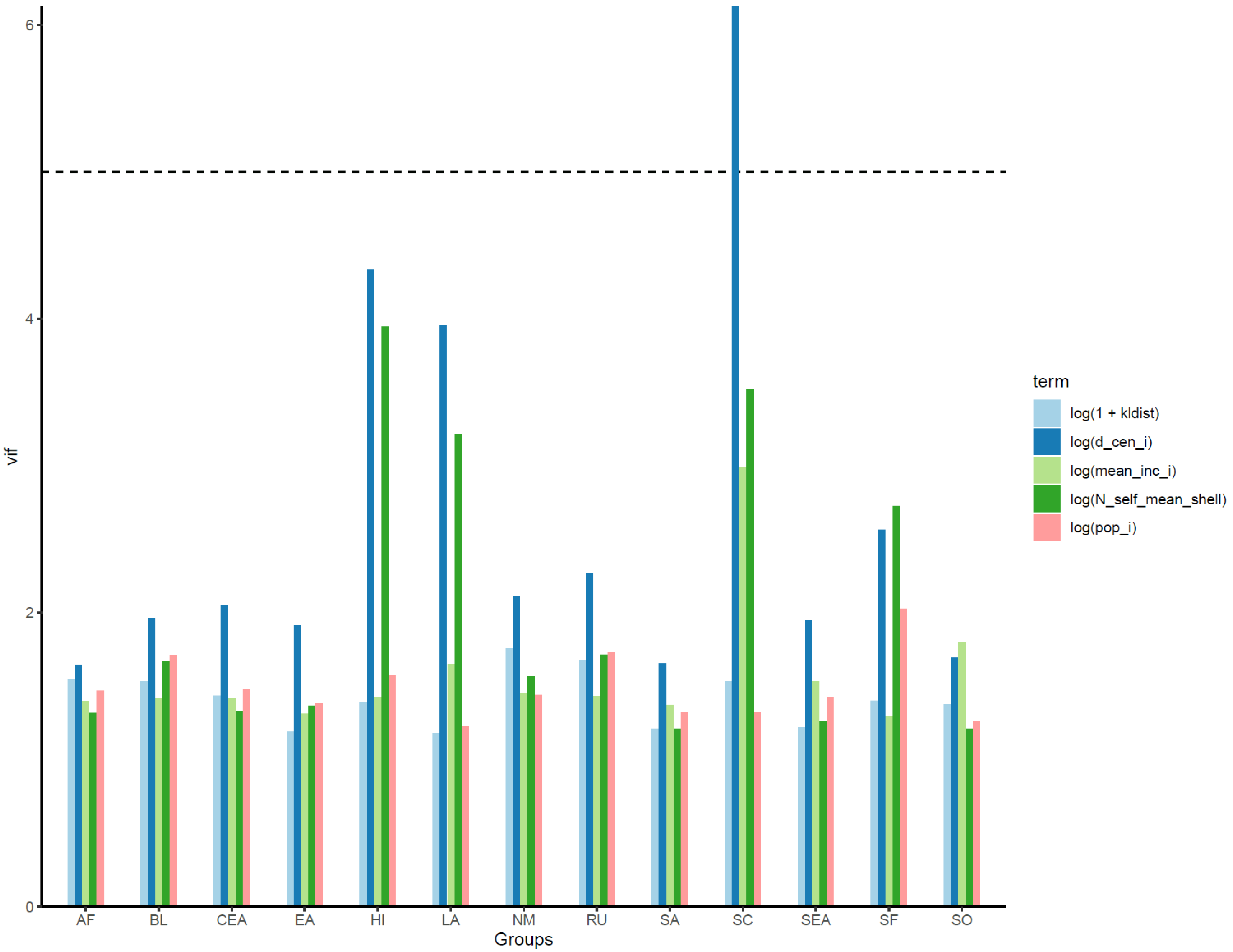}
\caption{Variance inflation factors (vifs) for different explanatory variables for the spatial full model in Eq.~3.  
}
\label{fig-appendix-4}
\end{figure}

\begin{figure}[h]
\centering
\includegraphics[width=0.7\textwidth]{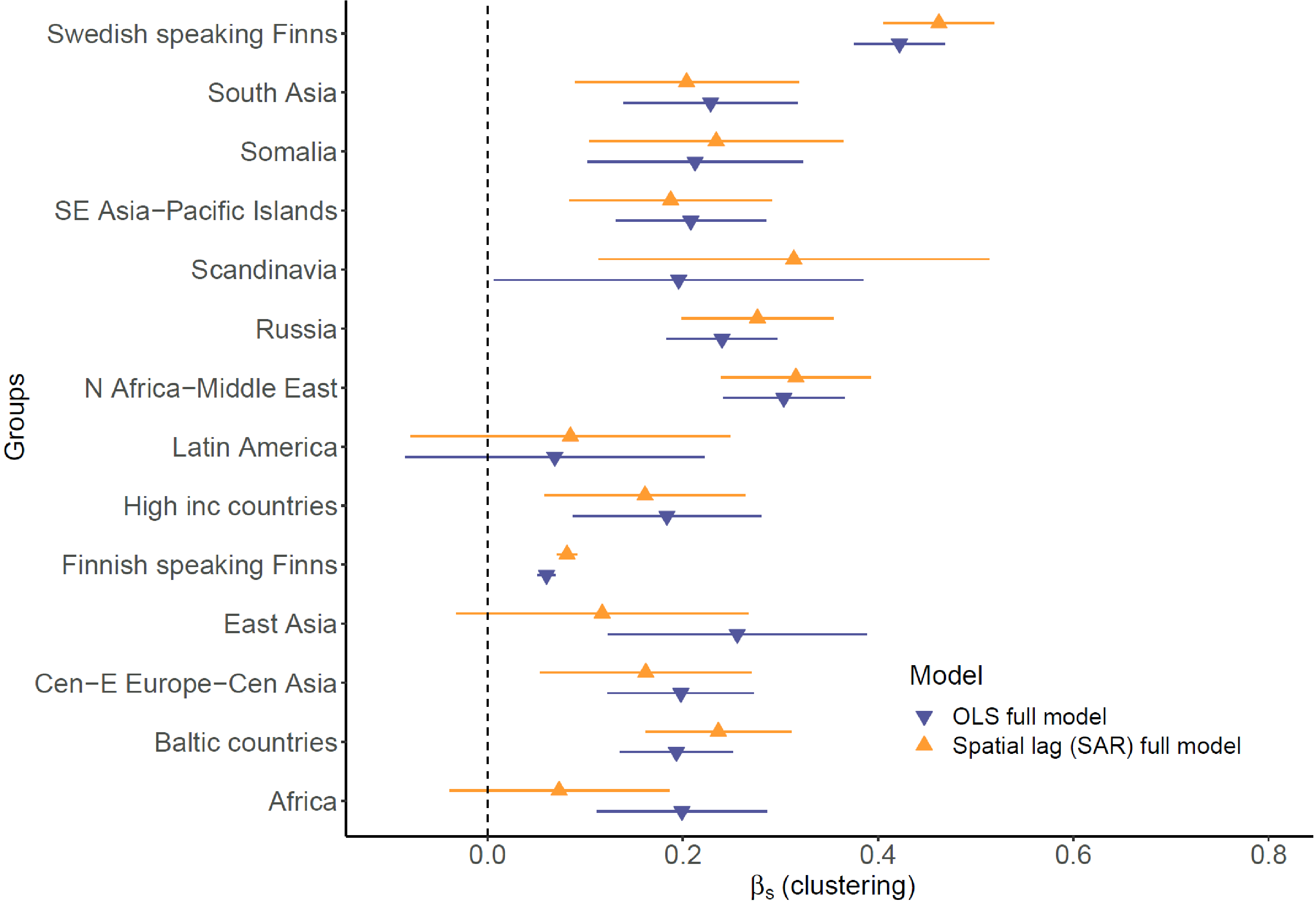}
\caption{Comparison between the ordinary least square (OLS) estimates and spatial autoregressive (SAR) lag model estimates for clustering. 
}
\label{fig-appendix-16}
\end{figure}

\begingroup
\squeezetable
\setlength{\tabcolsep}{18pt}
\renewcommand{\arraystretch}{1.5}
\begin{table}[h]
\scriptsize
\caption {The fit indices corresponding to nested versions of the spatial model using hierarchical linear modelling. The full data set is used except the group of native Finnish speakers (FF). For the models we assume that only the slope coefficients have group level variation and the error terms are normally distributed with zero means and zero covariances.  The following indices are obtained by using a maximum likelihood estimation.} 
\begin{ruledtabular}
\begin{tabular}{lrrr}
\multirow{2}{*}{\textbf{Fit indices}} & \multicolumn{1}{c}{\textbf{Null model}} & \multicolumn{1}{c}{\textbf{Base model}} & \multicolumn{1}{c}{\textbf{Full model}} \\ 
 & \multicolumn{1}{c}{(Eq.~1)} & \multicolumn{1}{c}{(Eq.~2)} &   \multicolumn{1}{c}{(Eq.~3)}\\ 
   \hline \hline

 {\scriptsize Observations} & {\scriptsize 4430} & {\scriptsize 4430} & {\scriptsize 4430} \\

 {\scriptsize Groups} & {\scriptsize 13} & {\scriptsize 13} & {\scriptsize 13} \\

 {\scriptsize $\varDelta$(AIC)} &  & {\scriptsize -1124} & {\scriptsize -1302}\\

 {\scriptsize $\varDelta$(BIC)} &  & {\scriptsize -1117} & {\scriptsize -1283}\\
 
 {\scriptsize $\varDelta$(model df)} &  & {\scriptsize 1}& {\scriptsize 3}\\

 {\scriptsize $\varDelta$(-2$\times$Log Likelihood)} & {\scriptsize } & {\scriptsize 563 ($p<0.001$)} & {\scriptsize 654 ($p<0.001$)}\\

 {\scriptsize pseudo-$R^2$} & {\scriptsize 0.49} & {\scriptsize 0.61}& {\scriptsize 0.72}

\end{tabular}
\end{ruledtabular}
\label{table-s3}
\end{table}
\endgroup

\begin{figure}[h]
\centering
\includegraphics[width=0.7\textwidth]{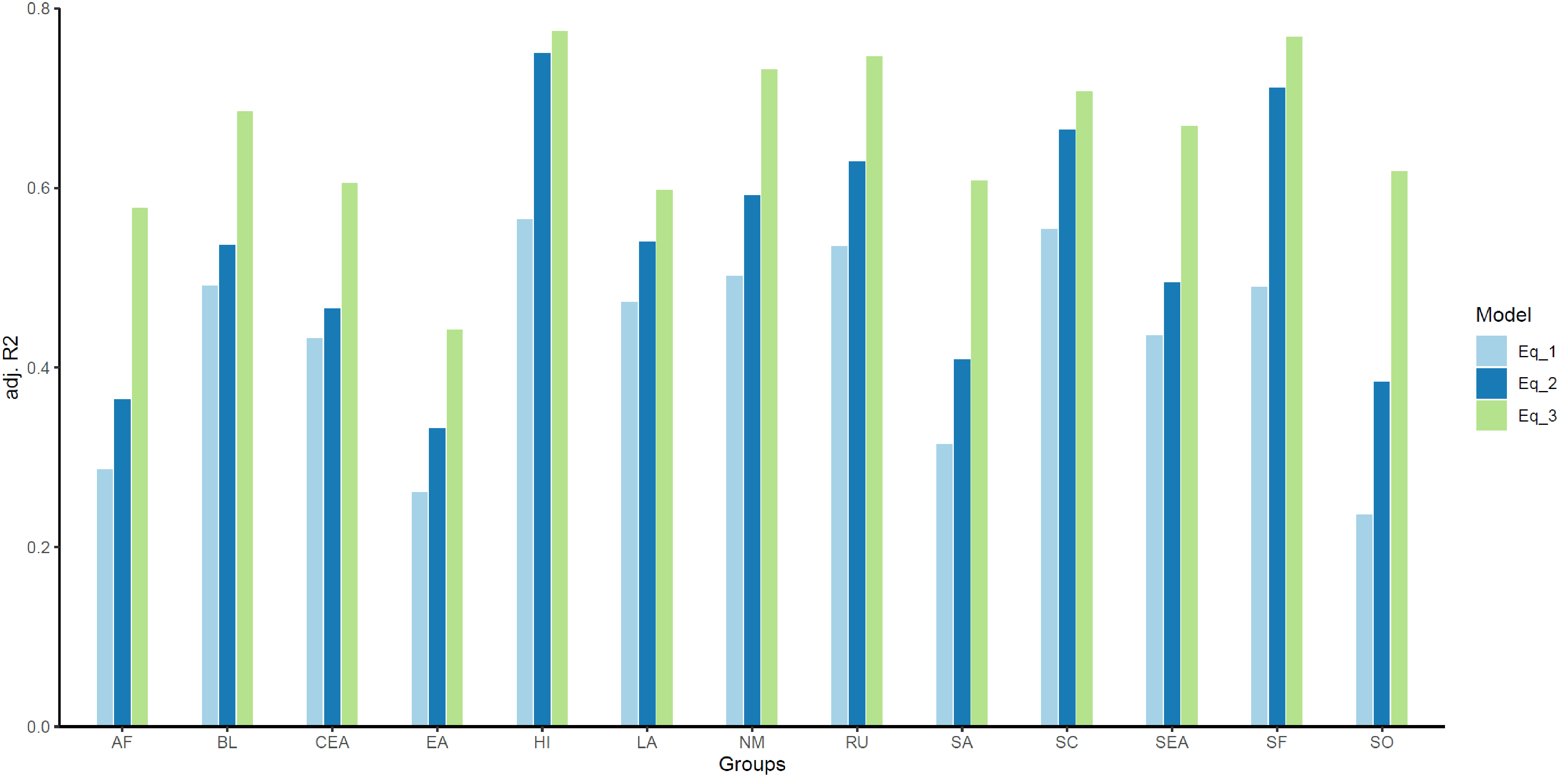}
\caption{Adjusted-$R^2$ values for the nested spatial models (null model, base model and full model).  
}
\label{fig-appendix-5}
\end{figure}

\begin{figure}[h]
\centering
\includegraphics[width=0.95\textwidth]{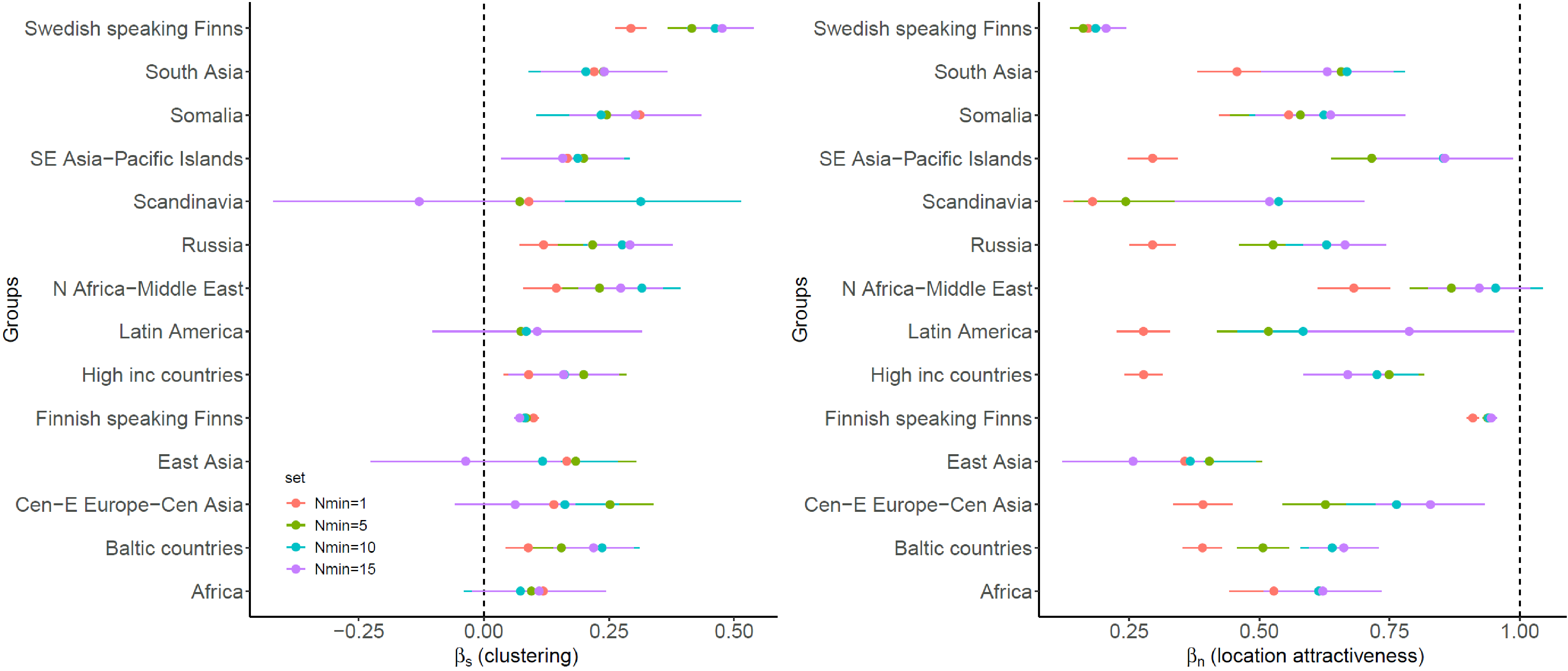}
\caption{The effect of data filtering on coefficients of the spatial model is shown. The spatial dataset for the analysis is derived from the original data after applying  a filtering $n_{i,g}\geq N_\text{min}$ and $n_i>n_{i,g}$ where $n_{i,g}$ and $n_i$ are the group population and total population in an area unit $i$, respectively. The reported coefficients are obtained with the dataset summarised in Table S2 which corresponds to $N_\text{min}=10$. In the above plots we show the how the coefficients $\beta_s$ and $\beta_n$ change when $N_\text{min}$ is changed for the spatial full model. The results shown are from spatial lag (SAR) models where filtering may have stronger effects compared to OLS. In general, for SAR estimation all area units that do not have neighbourhoods containing finite values of $n_{i,g}$ would be removed to preserve a square structure for the weight matrix. Therefore, a filtered dataset is subjected to further shrinkage in SAR.}
\label{fig-appendix-13}
\end{figure}

\begin{figure}[h]
\centering
\includegraphics[width=0.85\textwidth]{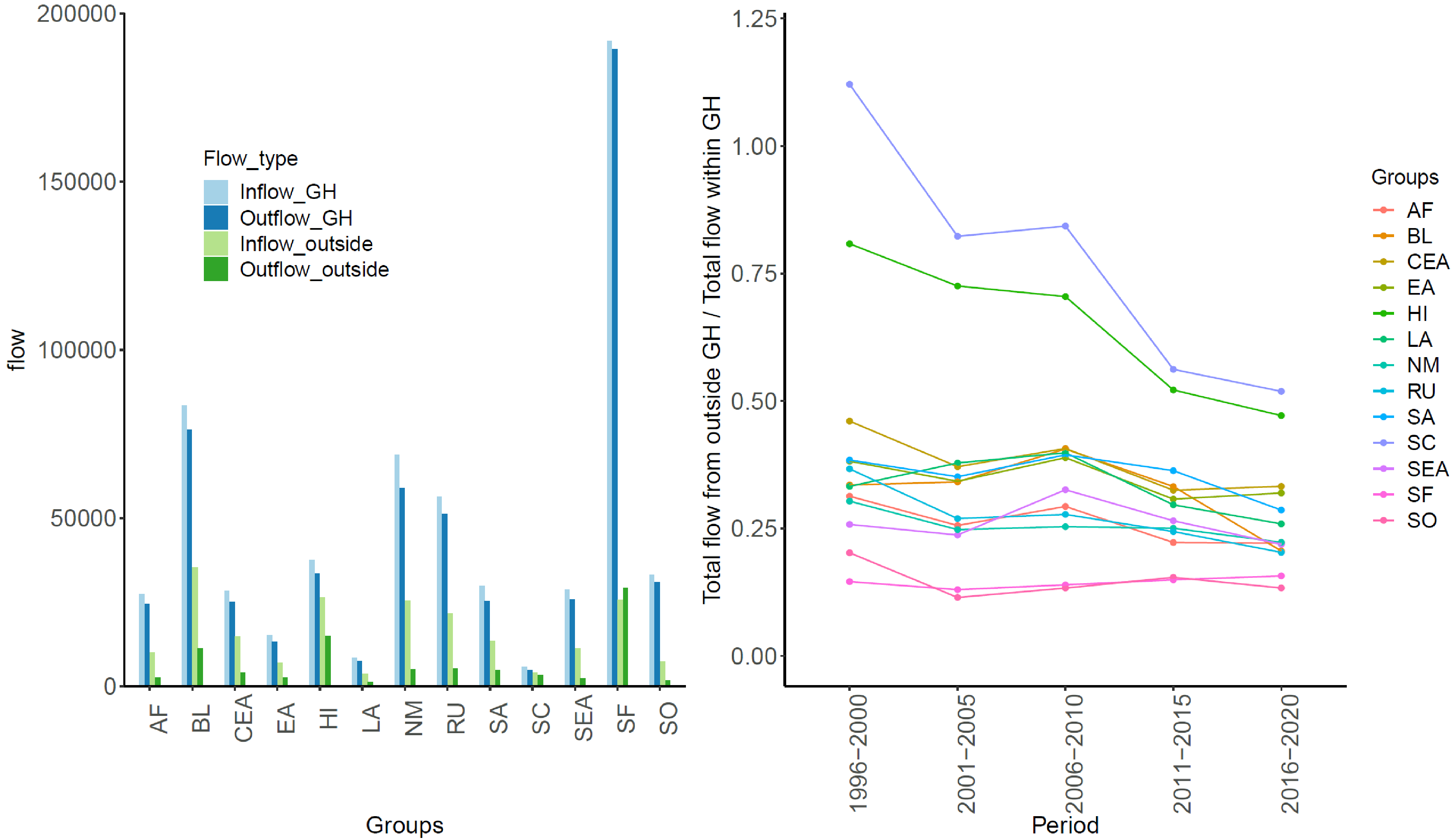}
\caption{The plots demonstrate that frequency intracity moves highly surpasses those resulting from immigration, emigration, and migration from or to parts of Finland outside Greater Helsinki (GH). (Left) Shows the total flux aggregated over all area units during 1996-2020 that was either inward directed (inflow) or outward directed (outflow). Filtering the migration dataset by a $N_\text{min}$ (see Fig.~\ref{fig-appendix-13}) and  by non-zero values of explanatory variables (using logarithms) causes a slight mismatch of these quantities which ideally should be balanced. The flows from or to outside Greater Helsinki is also shown. Except for native Swedish speakers (SF) for all the other groups inflow is larger than the outflow. (Right) Ratio of magnitude of flows from or to outside the region to flows originating within the region. The ratios are well below unity for the majority of the period and majority of the groups.}
\label{fig-appendix-15}
\end{figure}

\begin{figure}[h]
\centering
\includegraphics[width=0.7\textwidth]{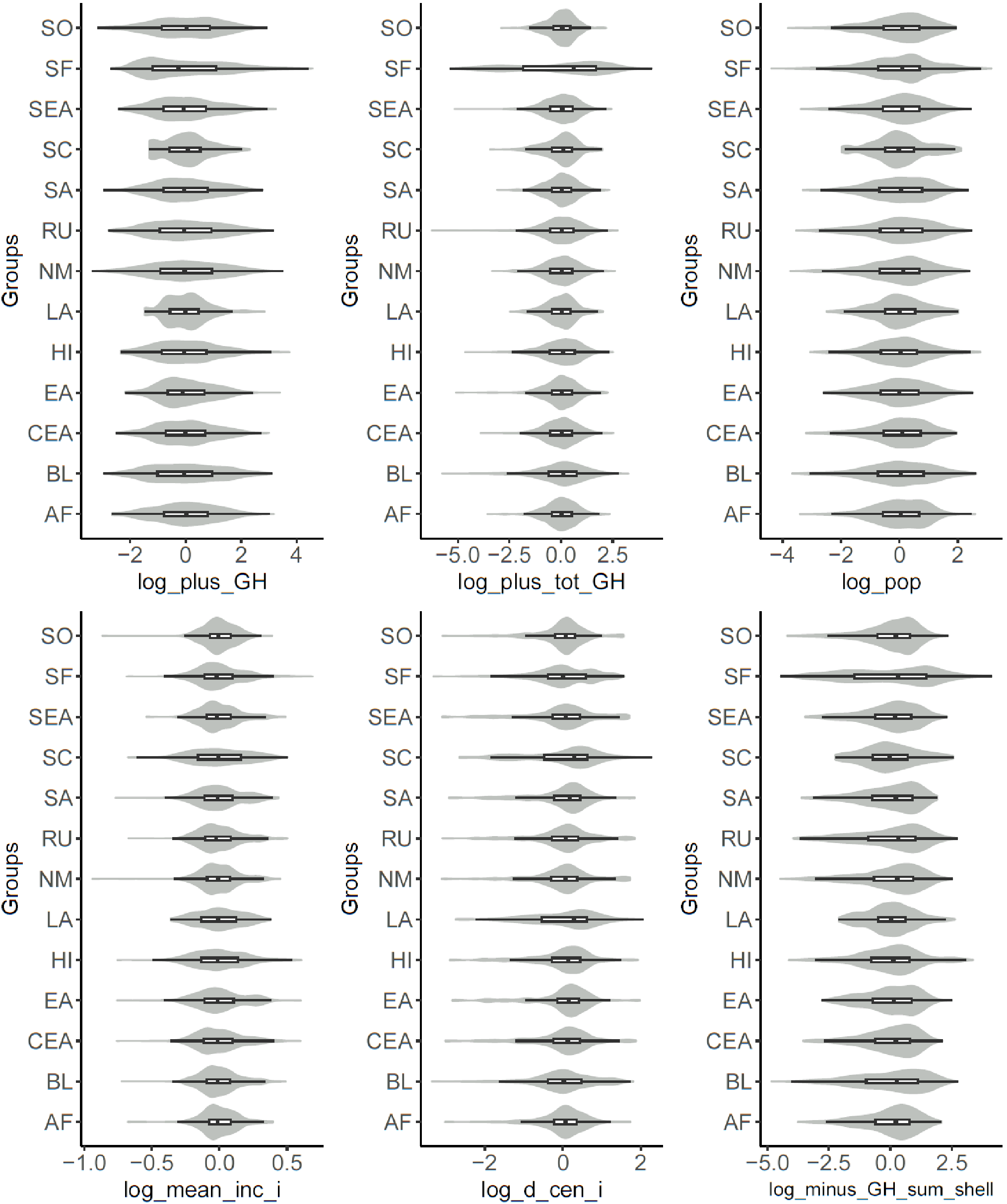}
\caption{Descriptive statistics for the variables in the migration model for inflow (clockwise from top-left):   $\log \mathcal{F}_{i,g}^\text{(in)}$, $\log \mathcal{F}_{i}^\text{(in)}$, $\log n_{i,g}$, $\log \widetilde{\mathcal{F}}_{i,g}^\text{(out)}$, $\log d_{\text{cen},i}$, $\log \langle m\rangle_{i}$. The variables $n_{i,g}$ and $\langle m\rangle_{i}$ correspond to the years 1995, 2000, 2005, 2010, and 2015. The other flow-related variables ($\mathcal{F}$) were aggregated over five year windows corresponding to the above years. The log transformed variables were mean-centred.}
\label{fig-appendix-9}
\end{figure}

\begin{figure}[h]
\centering
\includegraphics[width=0.7\textwidth]{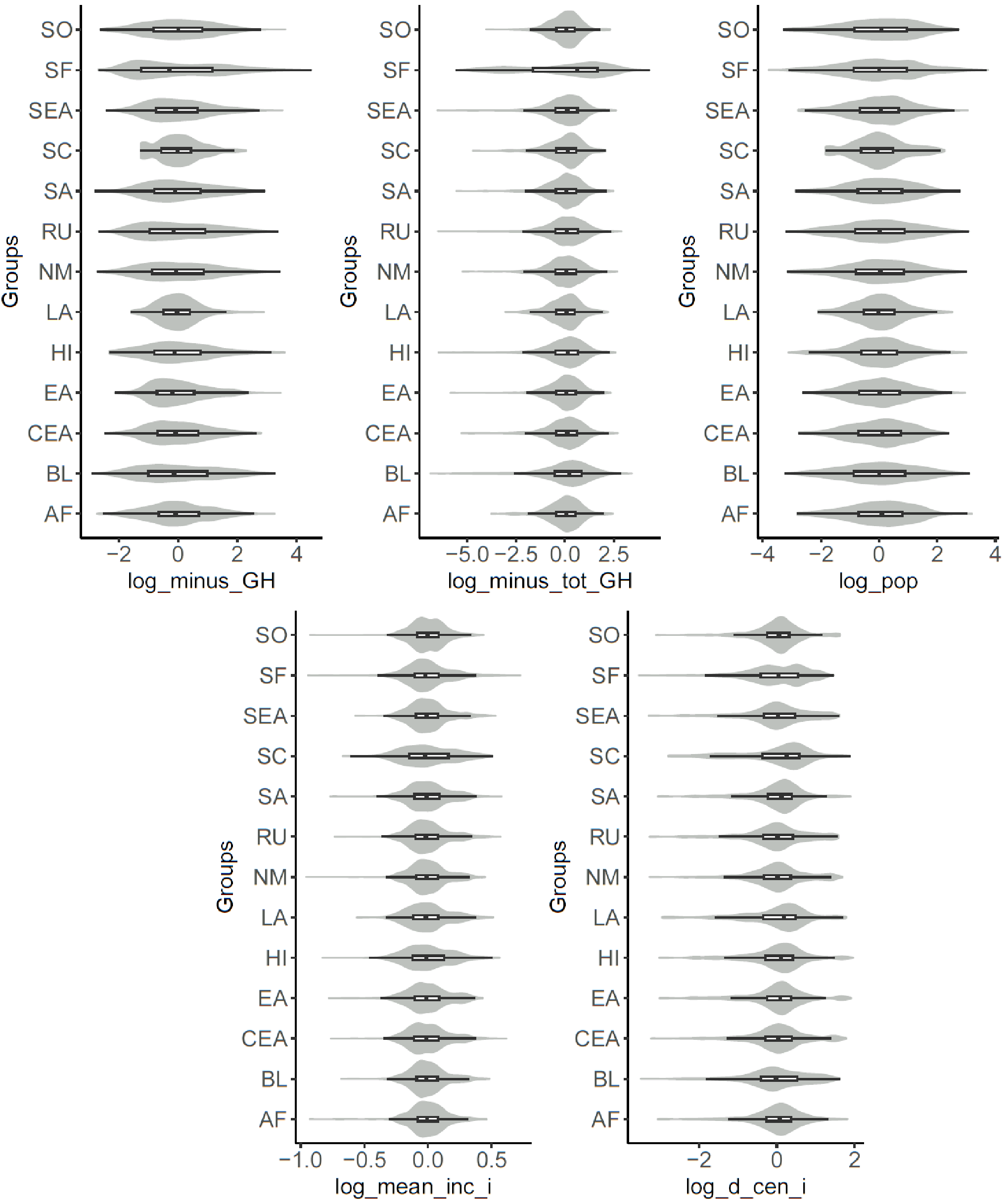}
\caption{Descriptive statistics for the variables in the migration model for outflow (clockwise from top-left):   $\log \mathcal{F}_{i,g}^\text{(out)}$, $\log \mathcal{F}_{i}^\text{(out)}$, $\log n_{i,g}$,  $\log d_{\text{cen},i}$, $\log \langle m\rangle_{i}$. The variables were obtained as described in Fig.~\ref{fig-appendix-9}.}
\label{fig-appendix-10}
\end{figure}

\begin{figure}[h]
\centering
\includegraphics[width=0.9\textwidth]{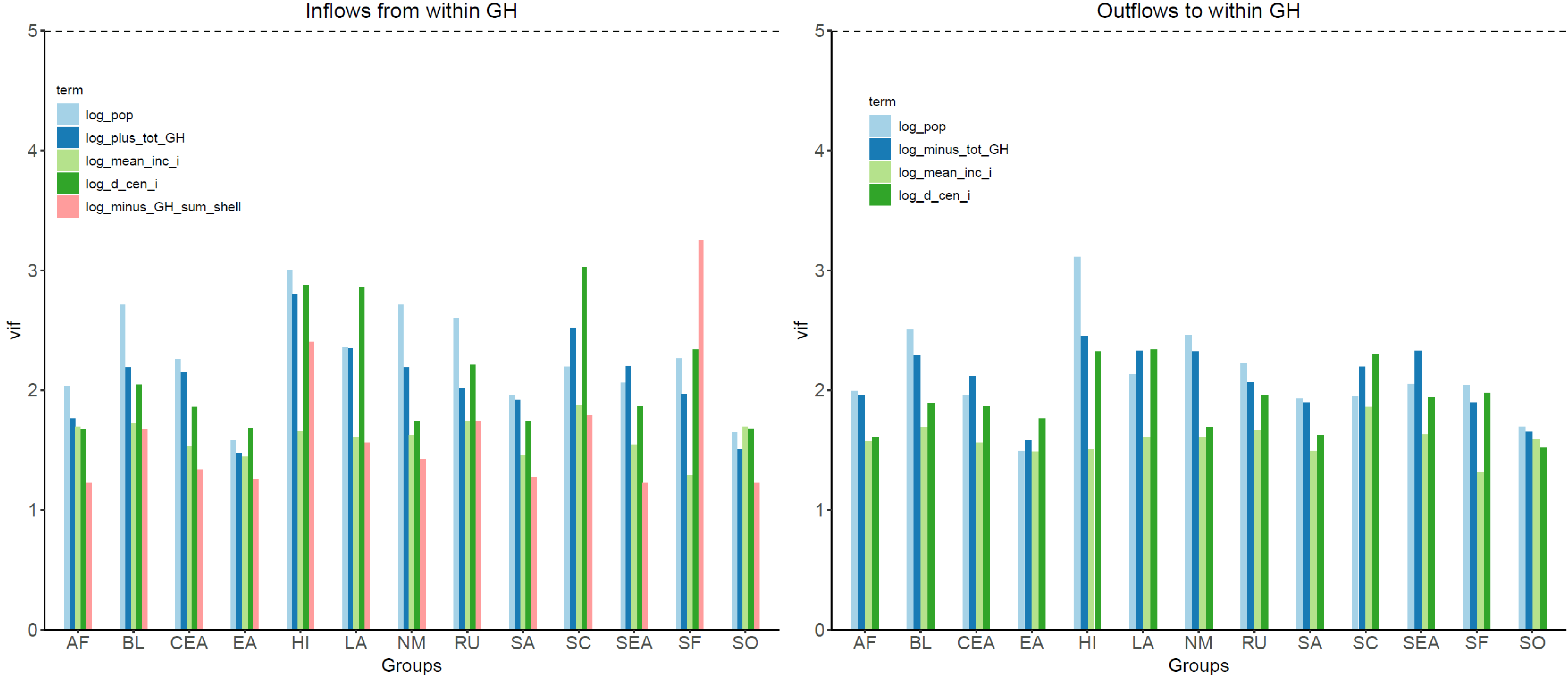}
\caption{Variance inflation factors (vifs) for different explanatory variables in the migration models.
}
\label{fig-appendix-7}
\end{figure}

\begin{figure}[h]
\centering
\includegraphics[width=0.9\textwidth]{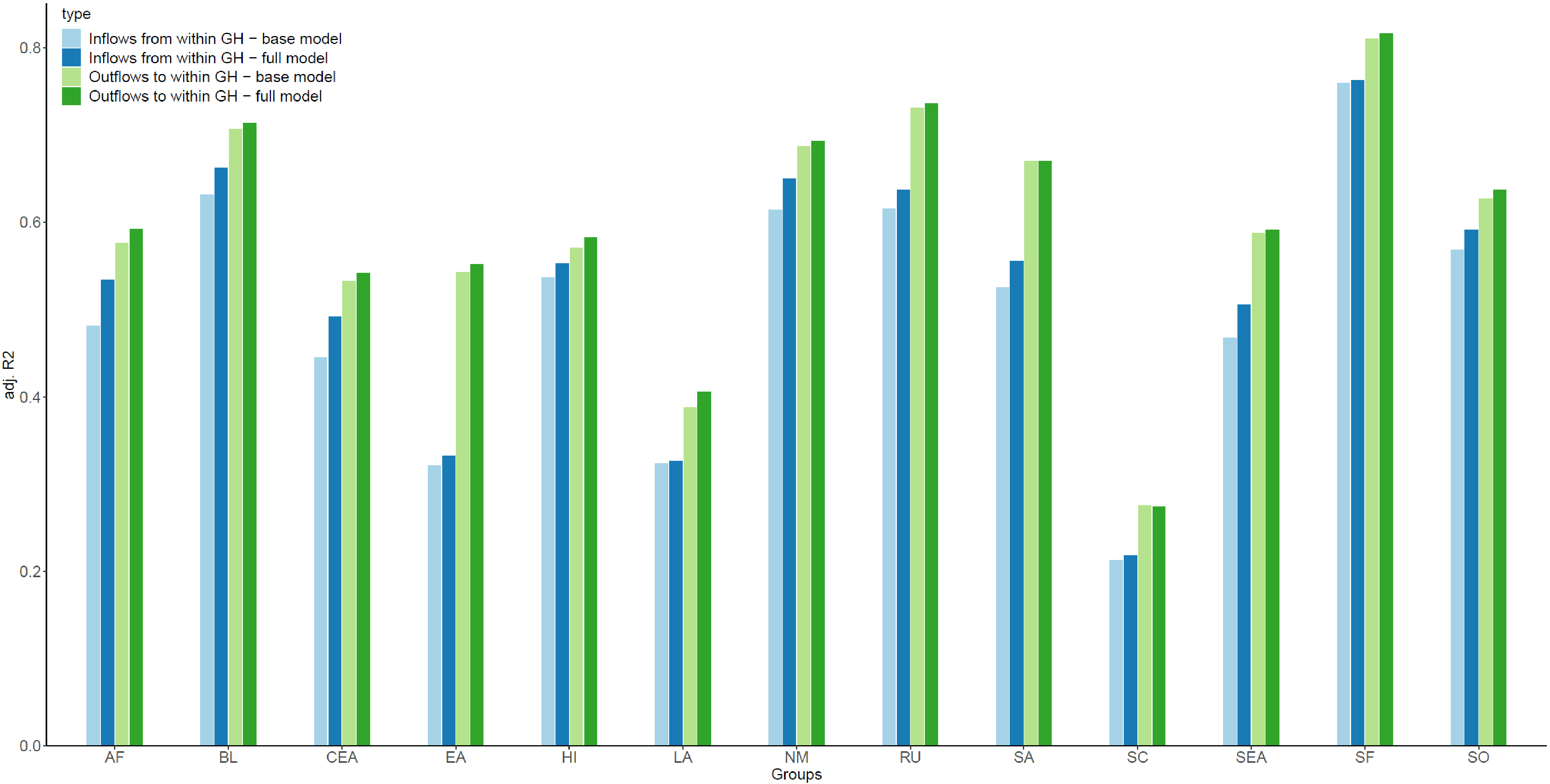}
\caption{Adjusted-$R^2$ values for the migration models.}
\label{fig-appendix-8}
\end{figure}

\begin{figure}[h]
\centering
\includegraphics[width=0.9\textwidth]{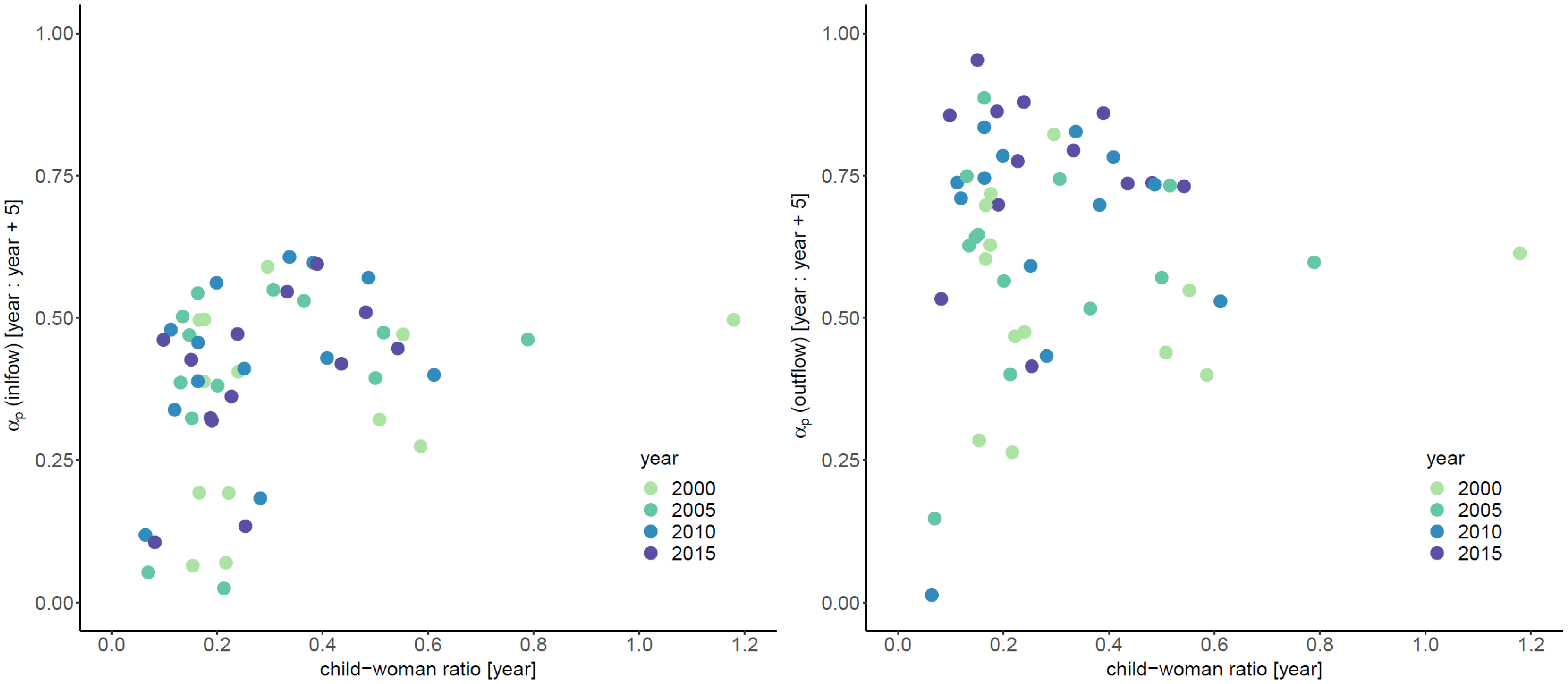}
\caption{The coefficients (exponent) of group population $\alpha_p ^\text{(in)}$ and $\alpha_p^\text{(out)}$ for the inflow and outflow models, respectively, are plotted against the child-woman ratio. Each data point represents the regression coefficient corresponding to a group and a specific period over which the flows were aggregated. While the CWR was calculated for the years ($\tau$) as shown in the legend, the flows corresponded to $[\tau,\tau+\Delta]$ with  $\Delta=5$. }
\label{fig-appendix-17}
\end{figure}

\end{document}